\documentclass{aa}  % for a paper on 2 columns  
\usepackage[table]{xcolor}
\usepackage{graphicx}
\usepackage{txfonts}
\usepackage{color}
\usepackage{multirow}
\usepackage{soul} %LK added to be able to strikethrough text

\begin{document}

   \title{A comparison of the active region upflow and core properties using simultaneous spectroscopic observations from IRIS and Hinode}

   \author{Krzysztof Barczynski \inst{1,2} \and 
          Louise Harra\inst{1,2} \and Lucia Kleint\inst{3,4} \and Brandon Panos\inst{4,3}\ and
          David H. Brooks\inst{5}
          }

   \institute{PMOD/WRC, Dorfstrasse 33, CH-7260 Davos Dorf, Switzerland\\
              \email{krzysztof.barczynski@pmodwrc.ch}
         \and
             ETH-Zurich, H\"onggerberg campus, HIT building, Z\"urich, Switzerland
      %  \and 
     %   Kiepenheuer Institut f{\"u}r Sonnenphysik (KIS), Sch{\"o}neckstrasse 6, D-79104 Freiburg, Germany
     \and
        University of Geneva, CUI, 1227 Carouge, Switzerland
      \and
        University of Applied Sciences and Arts Northwestern Switzerland, Bahnhofstrasse 6, 5210 Windisch, Switzerland
             \and
        College of Science, George Mason University, 4400 University Drive, Fairfax, VA 22030 USA
             }

  \titlerunning{A comp. of the AR upflow and core prop. using simultaneous spectroscopic obs. from IRIS and Hinode}
\authorrunning{Barczynski et al.}

  % \date{Received ...; accepted 07.04.2021}
    \date{Accepted: 07.04.2021}

% \abstract{}{}{}{}{} 
% 5 {} token are mandatory

  \abstract
   {The origin of the slow solar wind is still an open issue. It has been suggested that upflows at the edge of active regions are a possible source of the plasma outflow, and therefore contribute to the slow solar wind.}
   {Here, we investigated the origin and morphology of the upflow regions and compared the upflow region and the active region core properties.}
   {We studied how the plasma properties (flux, Doppler velocity, and non-thermal velocity) change throughout the solar atmosphere, from the chromosphere via the transition region to the corona in the upflow region and the core of an active region. We studied limb-to-limb observations of the active region (NOAA 12687) obtained between 14th and 25th November 2017. We analysed spectroscopic data simultaneously obtained from IRIS and Hinode/EIS in six emission lines (\ion{Mg}{ii} 2796.4\AA,  \ion{C}{ii} 1335.71\AA, \ion{Si}{iv} 1393.76\AA, \ion{Fe}{xii} 195.12\AA, \ion{Fe}{xiii} 202.04\AA, and \ion{Fe}{xiv} 270.52\AA~and 274.20\AA). We studied the mutual relationships between the plasma properties for each emission line, as well as comparing the plasma properties between the neighbouring formation temperature lines. To find the most characteristic spectra, we classified the spectra in each wavelength using the machine learning technique k-means. }
   {We found that in the upflow region the Doppler velocities of the coronal lines are strongly correlated, but the transition region and coronal lines show no correlation.
   However, their fluxes are strongly correlated.
   The upflow region has lower density and lower temperature than the active region core.
   In the upflow region, the Doppler velocity and non-thermal velocity show a strong correlation in the coronal lines, but the correlation is not seen in the active region core.
   At the boundary between the upflow region and the active region core, the upflow region shows an increase in the coronal non-thermal velocity, the emission obtained from the DEM, and the domination of the redshifted regions in the chromosphere.}
   {The obtained results suggest that at least three parallel mechanisms  generate the plasma upflow: (1) the reconnection between closed loops and open magnetic field lines in the lower corona or upper chromosphere; (2) the reconnection between the chromospheric small-scale loops and open magnetic field; (3) the expansion of the magnetic field lines that allows the chromospheric plasma to escape to the solar corona.}


  \maketitle

%-------------------------------------------------------------------

\section{Introduction} \label{sec:intro}
The solar atmosphere continuously generates streams of particles - the solar wind that creates  space weather and the space environment in the Solar System.
These particles influence the Earth's ionosphere and magnetosphere and can have a negative impact on technology and astronauts' health. 
The solar wind velocity distribution presents two components of a different origin.
The source of the fast solar wind ($\approx$800 km/s at 1AU) is well established to be coronal holes, but the origin of the slow solar wind ($\approx$300 km/s at 1AU) is still an open issue.
The source of the slow solar wind has been considered to emanate from the coronal hole boundaries \citep{Wang1990}, helmet streamers \citep{Einaudi1999, Wang2000}, or the edges of the active region \citep{Kojima1999, Sakao2007, Harra2008}.
In this paper, we focus on the border between the active region and the coronal hole to investigate the origins of the slow solar wind.

Doppler velocity maps, obtained at coronal temperatures, always show a plasma upflow at edges of  active regions  \citep{Sakao2007, Doschek2007, Doschek2008, DelZanna2008, Hara2008, Harra2008}.
The upflow regions have been documented in all observations of active regions regardless of the size and complexity and stage of evolution.
\citet{Harra2017} have investigated an upflow region during 3 solar rotations, but \citet{Zangrilli2016} have found an upflow region that survived five solar rotations.
The plasma upflows have been considered as a source of the solar wind \citep{Sakao2007, Harra2008}.
However, the physical processes that generate the upflow regions are still not confirmed, and the following scenarios are considered: plasma circulation in the open coronal magnetic field funnels \citep{Marsch2008}, impulsive heating at the footpoints of AR loops \citep{Harra2008},  chromospheric evaporation due to reconnection driven by flux emergence and braiding by photospheric motions \citep{DelZanna2008}, chromospheric jets and spicules \citep{DePontieu2009, DePontieu2017}, Alfv\'en waves triggered by magnetic reconnection \citep{Wang2015}, heating events \citep{Nishizuka2011}, expansion of large-scale reconnecting loops \citep{Harra2008}, continual AR expansion \citep{Murray2010}, the reconnection between the closed-loop in the active region core and open magnetic field lines create strong pressure imbalances responsible for the plasma upflow \citep{DelZanna2011}, and the AR magnetic field configuration partially uncovered by the streamers, that allows plasma to easily escape \citep{vanDriel2012}.

%Outflow
Despite the upflow regions often being co-spatial with open magnetic field lines \citep{Sakao2007, Harra2008}, not all of these regions are necessarily related to the outflow region and not all of them are the source of the slow solar wind \citep{Edwards2016}.
The coronal outflows can originate from the places with a strong gradient of magnetic connectivity, called Quasi-Separatrix Layers (QSLs), where magnetic reconnection transforms the closed magnetic field loops into 'open' field or large-scale loops \citep{Baker2009, Baker2017}.
This magnetic field topology allows the coronal upflow material to escape and create the solar wind \citep{Sakao2007}.
Moreover, \citet{Brooks2020} showed that there are two components of the outflow emission in the active region: (1) a substantial contribution from expanded plasma that appears to have been expelled from closed loops in the active region core, (2) and a contribution from dynamic activity in the active region plage, with a composition signature that reflects solar photospheric abundances.
%

%Upflow
Several slow solar wind sources are considered; three main scenarios are presented below \citep{Abbo2016}.
(1) The expansion model assumes that quasi-static long-lived open magnetic field topology allows plasma particle escape \citep{Wang1990b, Wang1994}.
This solar wind speed depends on the flux tube expansion; the faster expansion implied the slower speed.
In the active region edges, the flux tube's expansion is faster than in the coronal hole; thus, the active region edges are a slow solar wind source. 
(2) Interchange reconnection between the open magnetic field and closed-field lines in the transition region eject particles onto open magnetic field lines creating the solar wind \citep{Fisk2001}.
(3) The weak magnetic field of the streamer (pseudostreamer) cusp allows the material in the underlying closed field-lines to escape as a result of the reconnection \citep{Einaudi1999, Rappazzo2005}.

%Pa 3. Spectroscopy + non-thermal line broadening
The origin of the coronal upflow region and its relation to the solar atmosphere's underlying layer is also an open question.

One important characteristic of the AR upflows is the blue-ward asymmetries of coronal line profiles \citep{Harra2008}.
Intensive investigations have been performed to understand these asymmetries.
The asymmetries also provide important clues on the generation mechanisms of the upflows \citep{Tian2011, McIntosh2012}.
However, the mechanisms of the upflow creation and location where in the solar atmosphere these mechanisms occur still remain undiscovered.
In this work, we concentrate on the processes responsible for the upflow region creation and evolution. 
To address this, we study the evolution of the plasma properties from the chromosphere through the transition region to the corona.
For plasma diagnostics, we use spectroscopic data obtained simultaneously by the Interface Region Imaging Spectrograph \citep[IRIS;][]{DePontieu2014} and Hinode EUV Imaging Spectrometer \citep[Hinode/EIS;][]{Culhane2007SoPh}.
These observations allow us to analyse the plasma motions (Doppler velocity), and the excess line width above the thermal width which is known as the  non-thermal velocity.  
Many scenarios have been considered to explain the non-thermal line broadening: propagation of acoustic or MHD waves \citep{Boland1975, Coyner2011}, the small-scale magnetic reconnection events -nanoflares \citep{Parker1988, Patsourakos2006}, unresolved laminar flow along with different size loops \citep{Athay1991} and superposition of two or more velocity components along the line of sight (the multiplicity of line-of-sight velocities) \citep{Chen2011, Tian2011}.
%

%Par.4 - what's unknown?
Although many papers discuss the dependencies between the intensity, Doppler shift, and non-thermal velocity in the upflow region, there is little information on how these parameters and their dependencies change with the line formation temperatures from the chromosphere via transition region to the solar corona.
Several authors have investigated the temperature dependence of the line parameters, but mainly focus on a narrow temperature range.
For instance, \citet{DelZanna2008, Tripathi2009, Warren2011}) have performed a detailed study of the Doppler velocities as a function of line formation temperature.
For a broader view, we analyse the evolution of the plasma properties in the upflow region in a wide range of temperatures that covers the chromosphere, the transition region, and the corona through the analysis of  simultaneous observations obtained by IRIS and Hinode (Sect.~\ref{sec:data_analysis}).
We use maps of plasma parameters to study their spatial distribution and relationships between each other (Sect.~\ref{sec:results_par}, ~\ref{sec:nth_velocity}).
Finally, we classify the most characteristic spectra in the active region core and upflow region using an artificial intelligence method to discuss differences observed (Sect.~\ref{sec:spectra_class}).
In Sect.~\ref{sec:discussion}, we discuss the relationships between the plasma parameters observed in the different spectral lines and outline the main conclusion and suggestion for further analysis.
The methods presented in this work can be used to study the high-resolution spectroscopic data \citep[SPICE;][]{Spice2020} from the Solar Orbiter mission \citep{Muller2013, Muller2020}.

\section{Data analysis} \label{sec:data_analysis}

\subsection{Observation}\label{sec:observation}
\begin{figure*}[!h]
\centering
\includegraphics{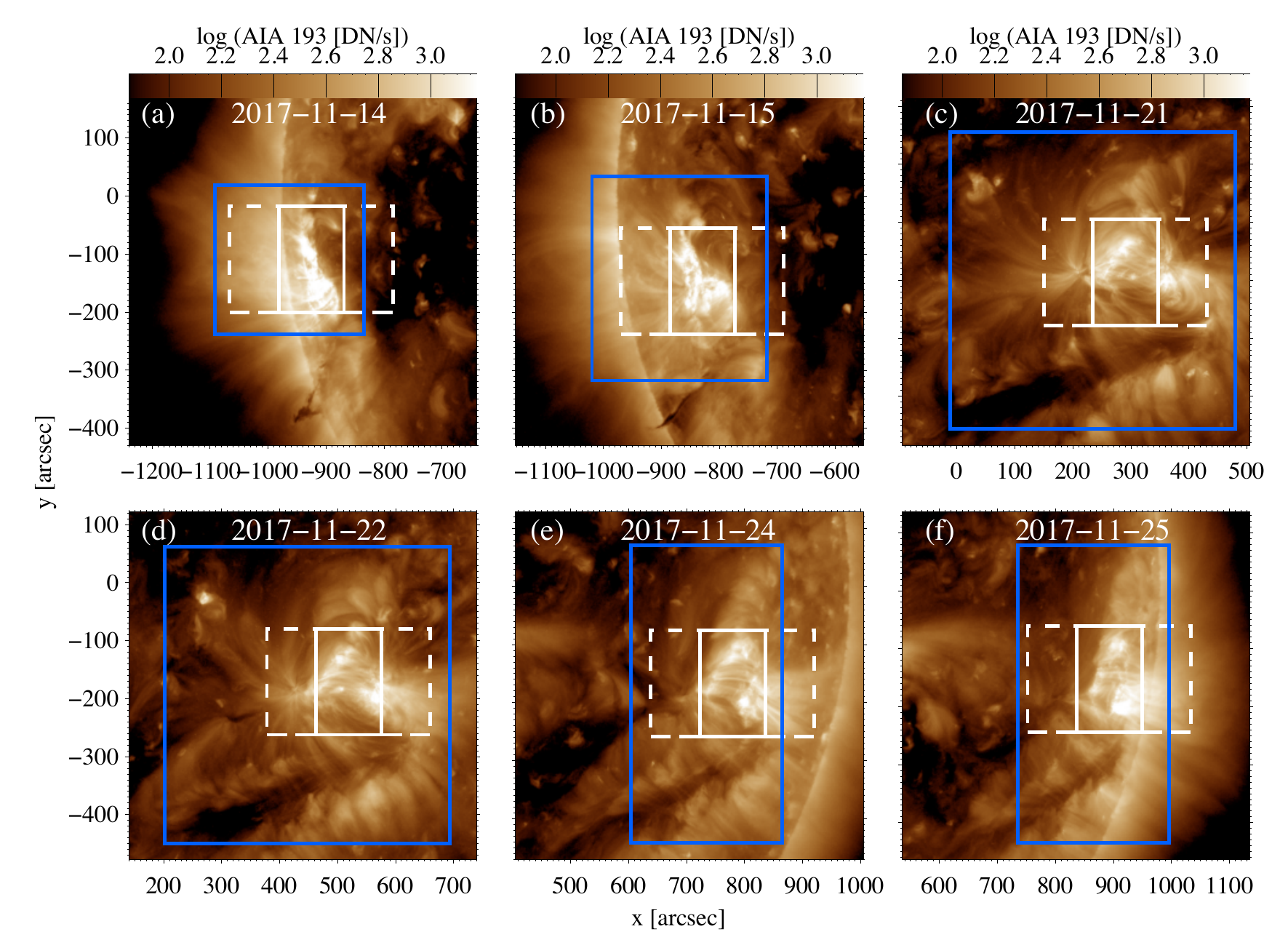}
\caption{Active region context and regions of interest with upflow and the active region core.
All panels show the emission of around 1.5MK observed in AIA193\AA~channel from 14 November 2017 to 25 November 2017.
The blue box indicates the Hinode/EIS full field-of-view.
The white dashed, and solid boxes highlight the IRIS slit-jaw and raster field-of-view, respectively.
The solid-white box of panel (b) is zoomed in Fig.~\ref{fig:tr_cor} and Fig.~\ref{fig:chromosphere}.}
\label{fig:aia193_overview}
    \end{figure*}

We analysed six datasets of the non-flaring NOAA Active Region AR12687 obtained between 14 and 25 November 2017.
For each observation, we selected a region of interest (ROI) in order to compare the properties of the upflow region and the active region core (Fig.~\ref{fig:aia193_overview}, Table~\ref{tab:obs_prop}).
We used spectroscopic data obtained simultaneously by IRIS  and Hinode/EIS to study the plasma properties.
These analyses are carried out for the temperature range from logT/K=3.8 to logT/K=6.3  (Table~\ref{tab:obs_overview}).
For reliable alignment between the Hinode and IRIS data, we used the images obtained by the Atmospheric Imaging Assembly \citep[SDO/AIA;][]{Lemen2012}. 
The SDO/AIA data are also used for the differential emission measure analysis (Sect.~\ref{sec:dem}).

% \usepackage{multirow}
\begin{table*}[h]
\caption{Properties of Hinode/EIS, IRIS, and SDO/AIA observations. Hinode/EIS acquired data with the 1"- 2" slit widths, 87-151 slit steps, and slit cadences of 31.9-101.8 s. IRIS collected data with the 0.35" slit width, 320 slit steps, and slit cadences of 9.2-9.6 s.}\label{tab:obs_prop}
\resizebox{\textwidth}{!}{\begin{tabular}{llllllll}
\hline\hline
Instrument              & Obs. details\textbackslash{}data & 14 Nov 2017     & 15 Nov 2017     & 21 Nov 2017     & 22 Nov 2017     & 24 Nov 2017     & 25 Nov 2017    \\\hline
\multirow{2}{*}{Hinode} & Time [hh:mm]                                   & 14:18 -- 16:29 & 12:35 -- 16:50 & 11:38 -- 12:43 & 12:47 -- 13:52 & 03:17 -- 04:17 & 03:52 -- 04:52 \\
                        & FOV(x,y)[arcsec]                                     & 258"x256"      & 302"x352"      & 491"x512"      & 491"x512"      & 260"x512"      & 260"x512" \\\hline
\multirow{2}{*}{IRIS}   & Time [hh:mm]                                    & 15:32 -- 16:31 & 14:16 -- 15:08 & 11:37 -- 12:26 & 11:59 -- 12:49 & 02:57 -- 03:47 & 03:19 -- 04:09 \\
                        & FOV(x,y)[arcsec]                                     & 112"x175"      & 112"x174"      & 112"x175"      & 112"x175"      & 112"x175"      & 112"x175"\\\hline
\end{tabular}}
\end{table*}

\begin{table}[htbp]
\caption{Spectral lines or passbands of Hinode/EIS, IRIS and SDO/AIA used in our study and used in our analysis.}\label{tab:obs_overview}
\centering
\begin{tabular}{@{}llllll@{}}
\hline\hline
Line/channel     & $\lambda$[\AA] & logT/K & Instr. & Usage &  \\ \hline
\ion{Fe}{xiv}   & 270.52; 274.20     & 6.3         & EIS & analysis       & \\
\ion{Fe}{xiii}  & 202.04     & 6.2         & EIS & analysis       &  \\
\ion{Fe}{xii}   & 195.12     & 6.0         & EIS & analysis       &  \\
\ion{Si}{iv}    & 1393.76    & 4.9         & IRIS & analysis       &  \\
\ion{C}{ii}     & 1335.71    & 4.6         & IRIS & analysis       &  \\
\ion{Mg}{ii} k3 & 2796.4     & 3.9         & IRIS & analysis       &  \\
\ion{Mg}{ii} k2 & 2796.6     &3.8         & IRIS & analysis       &   \\ \hline
\ion{Ca}{xvii}  & 192.858     & 6.7        & EIS & DEM        &  \\
\ion{Fe}{xvi}  & 262.984     & 6.4        & EIS & DEM        &  \\
\ion{Fe}{xii}  & 193.509     & 6.2        & EIS & DEM        &  \\
\ion{Fe}{xii}  & 186.880     & 6.2        & EIS & DEM, dens.        &  \\
\ion{Fe}{xi}  & 180.401     & 6.0        & EIS & DEM        &  \\
\ion{Fe}{x}  & 257.262     & 6.0        & EIS & DEM        &  \\
\ion{Fe}{x}  & 184.536     & 6.0        & EIS & DEM        &  \\
\ion{Fe}{viii}  & 185.213    & 5.8        & EIS & DEM        &  \\
\ion{He}{ii}  & 256.312     & 4.7        & EIS & DEM        &  \\\hline
AIA 94  & 94     & 6.8        & AIA & align., DEM        &  \\
AIA 335  & 335     & 6.4        & AIA & align., DEM        &  \\
AIA 211  & 211     & 6.3        & AIA & align., DEM        &  \\
AIA 193  & 193     & 6.1        & AIA & align., DEM        &  \\
AIA 171  & 171     & 5.8        & AIA & align., DEM       &  \\
AIA 131  & 131     & 5.6        & AIA & align., DEM       &  \\\hline
SJI 1400  & 1400     & 3.7-5.2         & IRIS & alignment       &  \\
\hline                      
\end{tabular}
\end{table}

\subsection{Hinode EIS} \label{sec:inst_eis}
%Intro
The EUV Imaging Spectrometer \citep[EIS;][]{Culhane2007SoPh} is a scanning slit spectrometer on-board Hinode.
The Hinode/EIS observes the upper transition region and the solar corona in two wavebands: 170-210\AA\ and 250-290\AA\ with a spectral resolution 0.0223\AA\ per pixel.
The Hinode/EIS can scan the field of view up to 9.3$\times$8.5 arcmin with the narrowest slit and a spatial scale of 1 arcsec per pixel.

%Data down. + calibration
We downloaded the level-0 data from the Hinode/EIS archive (\url{http://solarb.mssl.ucl.ac.uk/SolarB/SearchArchive.jsp}) and focused on the strong coronal emission lines \ion{Fe}{xii} (195.12\AA), \ion{Fe}{xiii} (202.04\AA), and \ion{Fe}{xiv} (270.52\AA, 274.20\AA).
Additionally, we used nine other spectral lines obtained from EIS for the differential emission measure analysis (Table~\ref{tab:obs_overview}, Sect.~\ref{sec:dem}).
To calibrate Hinode/EIS data, we used a standard routine eis\_prep that removed the CCD dark current, cosmic rays pattern, hot and dusty pixels from the detector exposures.
This routine also provided the radiometric calibration from the data number (DN) to the physical unit (erg cm$^{-2}$ s$^{-1}$ sr$^{-1}$ \AA$^{-1}$).
Moreover, we applied the correction for the slit tilt and the orbital variation of line position \citep{Warren2014}.

%Data analysis
Using the eis\_auto\_fit routine (Solar SoftWare,  \url{https://hesperia.gsfc.nasa.gov/ssw/hinode/eis/idl/analysis/line_fitting/auto_fit/eis_auto_fit.pro}), we fitted a single Gaussian function for \ion{Fe}{xii}, \ion{Fe}{xiii}, and \ion{Fe}{xiv} spectral lines at each spatial pixel.
For the \ion{Fe}{xii} lines, the $\chi^2$ test and the velocity error shows that the single and double Gaussian fits present the same results for our ROIs, hence we chose to use a single Gaussian fit.
The fitted parameters were used to generate the maps of the peak intensity and Doppler velocity.
Then, we converted the line-of-sight Doppler velocity to the radial velocity using the direction cosine correction ($\mu=\cos{\theta}$).
The direction cosine correction is used to reduce the systematic component of the projection effect related to the solar rotation.
Using the SolarSoft routine eis\_width2velocity, we computed the non-thermal velocity.
This routine is based on the equation:
FWHM$^2=($FWHM$_{instr})^2+4ln(2)(\lambda\backslash{c})^{2}($V$_t+($V$_{nt})^2)$, where
FWHM -full width at half maximum, FWHM$_{instr}$-instrumental width, $\lambda$ -wavelength of the peak of the emission line, c-speed of light, V$_{t}$ - thermal velocity, and V$_{nt}$ - non-thermal velocity.

\subsection{IRIS} \label{sec:inst_iris}
%Intro
The Interface Region Imaging Spectrograph \citep[IRIS;][]{DePontieu2014} is a space-based multi-channel imaging-spectrograph that investigates the chromosphere and the transition region. 
IRIS provides spectroscopic raster data of the solar atmosphere in two far-ultraviolet channels (FUV) 1332-1358\AA~and 1390-1406\AA, and a near-ultraviolet channel (NUV) 2785-2835\AA~with a spectral sampling of 12.8 m\AA\,for FUV and 25.6 m\AA\, for NUV.
The field of view depends on the observation mode, usually up to 135"x175", but larger raster are possible by re-pointing the satellite.

%Data down. + calib.
We analysed level-2 data, downloaded from the IRIS database (\url{https://iris.lmsal.com/search/}).
These data have been corrected for flat field and geometrical distortion and the dark current has been subtracted. 
The automatic wavelength calibration was confirmed using the \ion{O}{i} 1335.60\AA\, line for FUVS detector, the \ion{S}{i} 1401.515\AA\, for the FUVL detector, and \ion{Ni}{i} 2799.474\AA\, line for NUV detector.
We focused on several lines of \ion{Mg}{ii}, \ion{C}{ii}, and \ion{Si}{iv} (Table~\ref{tab:obs_overview}).
For all IRIS wavebands analysed in this paper, the raster data were binned to the spatial resolution of the Hinode/EIS, but the spectral resolution was not modified.
This binning increased the signal to noise ratio.

%Data analysis
To obtain the intensity and the velocity of the \ion{Mg}{ii} peaks, we modified the iris\_get\_mg\_features\_lev2 routine to work with the spatially binned data and restricted the velocity range to $\pm$40 km s$^{-1}$.
The missing data of intensity and velocity (flagged NaN) were replaced with the interpolated values from the four nearest pixels.
For the \ion{C}{ii} and \ion{Si}{iv} spectral lines, we fitted a single Gaussian profile.
From the fitted parameters, we created the intensity and Doppler velocity maps.
In the last step, the line-of-sight Doppler velocity maps were converted to radial velocity map using the direction cosine correction $\mu=\cos{\theta}$. 
The $\chi^2$ test (IDL XSQ\_TEST procedure) was used to test the goodness of fit for the \ion{C}{ii} and \ion{Si}{iv} line for each spatial pixel.
We tested the hypothesis that the given observed spectral line intensities are an accurate approximation to the expected Gaussian intensity distribution and assumed hypothesis significance level at 0.05 .
For further analysis, we used only the fits that fulfilled this hypothesis.

%SJI images
The Slit Jaw Imager (SJI) onboard IRIS provides contextual images, showing the area surrounding the slit in four relatively wide band-passes. 
The two channels of 55\AA\, bandpass are centred at 1330\AA\ and 1400\AA, the other two channels of 4\AA\ bandpass are centred at 2796\AA\ and 2832\AA.
We used the auxiliary SJI 1400\AA~images for data alignment (Sect.~\ref{sec:data_align}).

\subsection{SDO: Atmospheric imaging assembly (AIA)} \label{sec:inst_aia}

%~\citep[SDO/AIA;][]{Lemen2012}
The Atmospheric Imaging Assembly \citep[SDO/AIA;][]{Lemen2012} provides a continuous monitoring of the full-solar disk in seven EUV and three UV channels that covers the solar atmosphere from the photosphere to the corona.  
The AIA data are obtained with a spatial scale of 0.6 arcsec per pixel and a temporal resolution of 12 sec for EUV channels and 24 sec for UV channels.

We used pre-processed SDO data provided by the Joint Science Operations Center (JSOC; \url{http://jsoc.stanford.edu}).
The pre-processing option AIA\_SCALE was applied.
Therefore, data have been corrected for proper plate scaling, shifts, and rotation (i.e. equivalent to level-1.5 data).

To show context and for the alignment between  Hinode/EIS and IRIS observations, we used the images from SDO/AIA 304\AA, 171\AA, and 193\AA~channels (Table~\ref{tab:obs_overview}). 
Figure~\ref{fig:aia193_overview} shows the active region's context in AIA 193\AA~ with the boxes representing the IRIS (white) and Hinode/EIS (blue) field of view.
The Hinode/EIS field of view (FOV) is larger than IRIS FOV hence our region of interest is the latter (solid white box in Fig.~\ref{fig:aia193_overview}).
The SDO/AIA images of 6 channels were used for differential emission measure analysis (Sect.~\ref{sec:dem}).

\subsection{Data alignment}\label{sec:data_align}
The high-precision alignment of the IRIS and Hinode/EIS data has a direct influence on a comparison of the plasma properties from different layers of the solar atmosphere.
Hinode/EIS observes significantly hotter spectral lines than IRIS, and hence in order to align accurately we used both SDO/AIA and IRIS SJI 1400 observations. 
To keep the same spatial scale of all observations, we interpolated the IRIS and SDO/AIA data to the Hinode/EIS spatial scale.
Using a cross-correlation method, we aligned the intensity maps of the lines with close formation temperature. 
Hence, the maps were aligned according to the following chain: \ion{Mg}{ii} k2 - \ion{Mg}{ii} k3 - \ion{C}{ii} - \ion{Si}{iv} -SJI1400 -AIA 304 -AIA 171 -AIA 193 - \ion{Fe}{xii} - \ion{Fe}{xiii} - \ion{Fe}{xiv}.
In this chain, the IRIS raster data had a rigid instrumental alignment, similar to the \ion{Fe}{xii} and \ion{Fe}{xiii} rasters from EIS.
The \ion{Fe}{xiv} line spectrum is recorded on a different CCD than \ion{Fe}{xii} and \ion{Fe}{xiii} lines and  the  offset between these two CCDs of Hinode/EIS is determined using the routine eis\_ccd\_offset.pro.
Finally, IRIS and Hinode data were aligned with the accuracy of the single Hinode/EIS spatial pixel.

\section{Results}\label{sec:results_par}
\subsection{Maps of intensity, velocity and non-thermal velocity}\label{sec:observation_maps}
%Intro

From the six spectroscopic observations presented in Fig.~\ref{fig:aia193_overview} (Table~\ref{sec:data_analysis}), we chose the most representative one - the observation obtained on 15 November 2017 to show detailed plots.
This observation clearly shows the upflow region and the active region core.
The same analysis  was carried out for all observations and we present  the statistical results for the rest of the observations.
%

\begin{figure*}[ht!]
\centering
\includegraphics[scale=0.85]{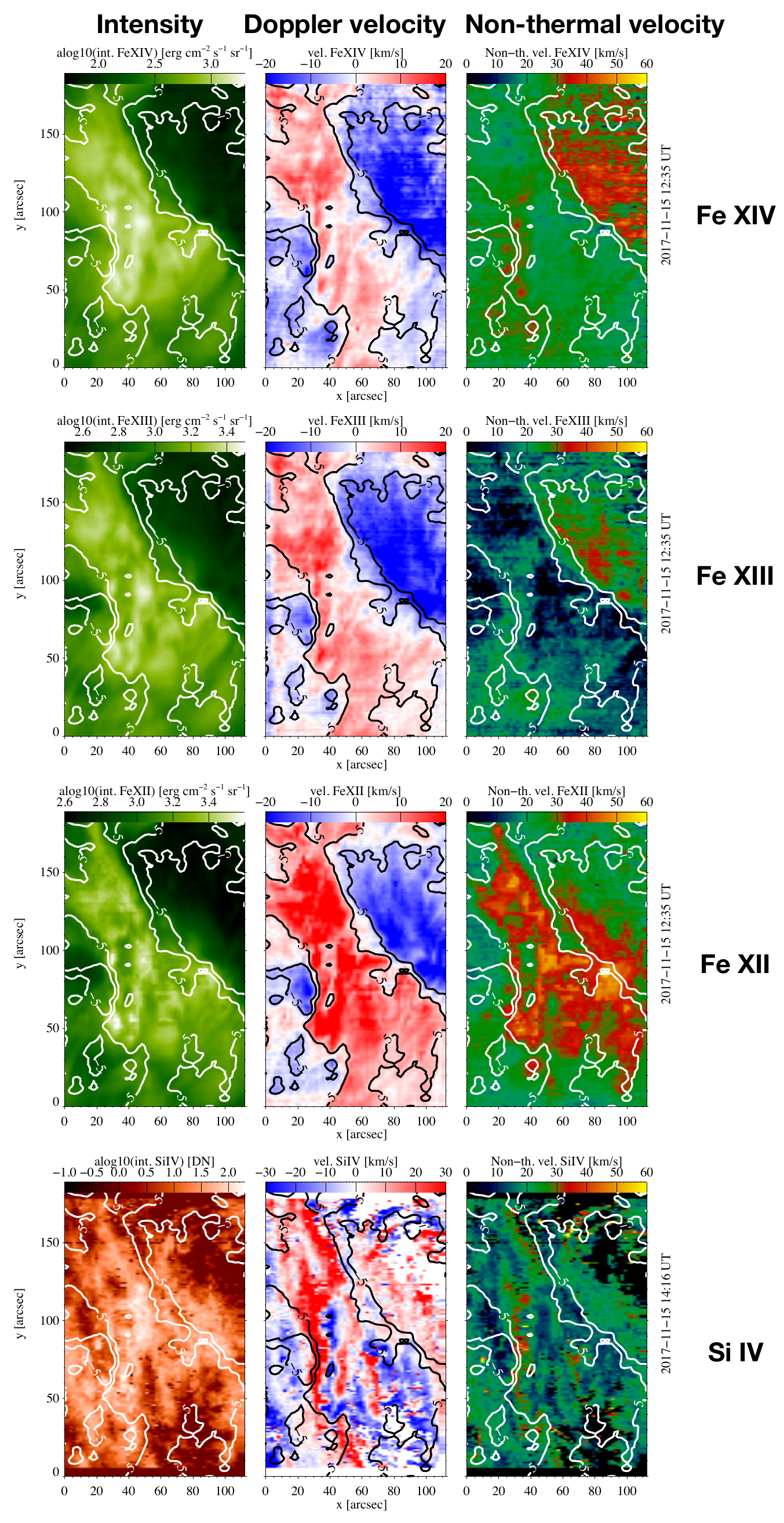}
\caption{
The corona and transition region of the active region core and upflow region observed with Hinode (first three upper row) and IRIS (bottom row) on 15 November 2017.
The panels show the intensity (left column), Doppler velocity (middle column), and non-thermal velocity (right column).
The rows are ordered with decreasing line formation temperatures from 2MK (\ion{Fe}{xiv} line, top row) to 0.08MK (\ion{Si}{iv}, bottom row).
To distinguish the active region core and upflow region, we added the contour line of the velocity in \ion{Fe}{xii} line of 5km s$^{-1}$ and -5km s$^{-1}$, respectively.}
\label{fig:tr_cor}
\end{figure*}

\begin{figure*}[ht!]
\centering
\includegraphics[scale=0.82]{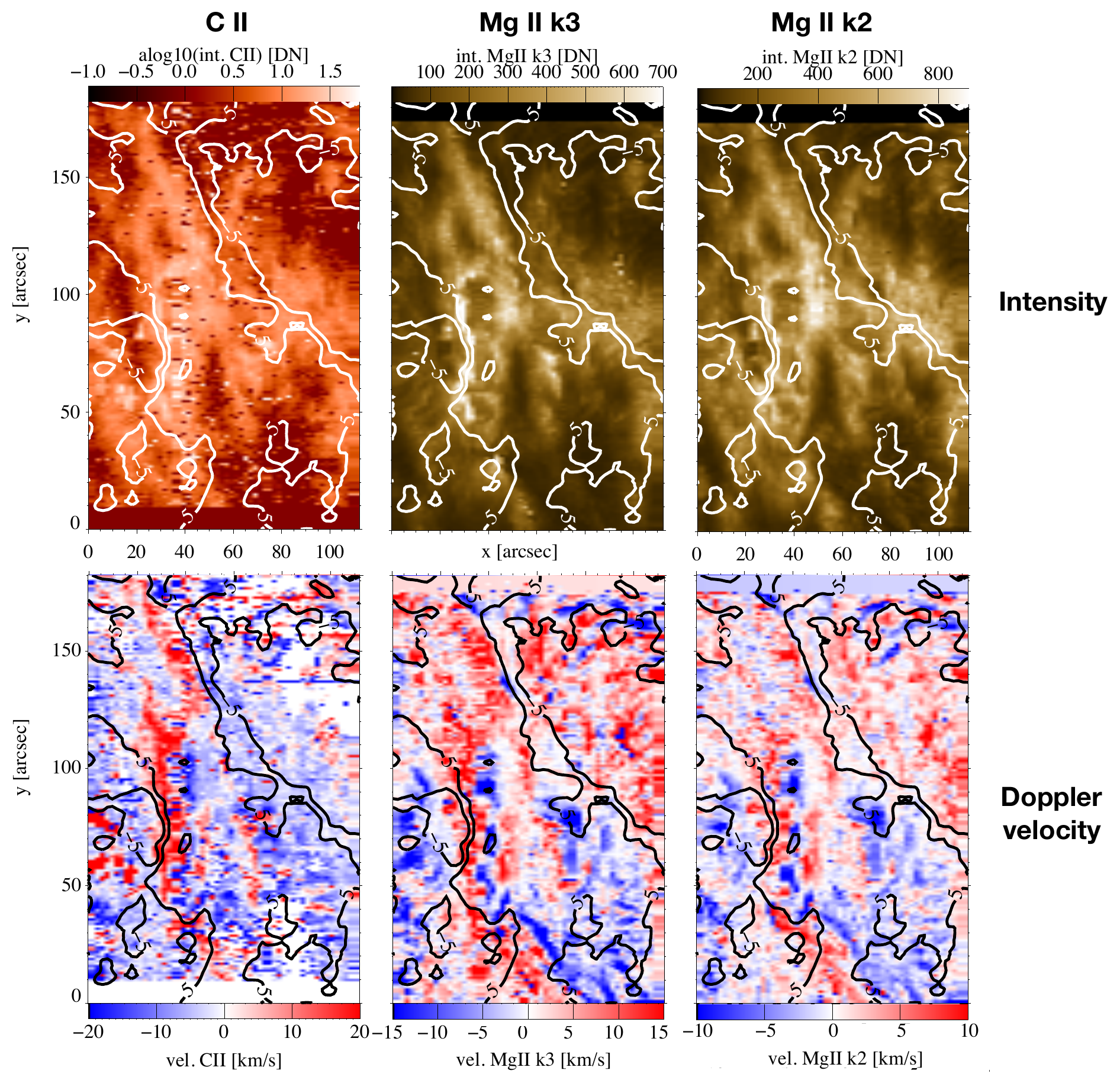}
\caption{
The transition region and chromosphere of the active region core and upflow region observed with IRIS on 15 November 2017.
The panels show the intensity (upper row) and Doppler velocity (bottom row).
The column is ordered with the decreasing line formation temperatures from 40 000 K (\ion{C}{ii} line, left column) to 6300 K (\ion{Mg}{ii} k2, right column).
To distinguish the active region core and upflow region, we added the contour line of the velocity in \ion{Fe}{xii} line of 5km/s and -5km/s, respectively (defined in Fig.~\ref{fig:tr_cor}).}
\label{fig:chromosphere}
\end{figure*}

%Upflow/AR core definition
Figures~\ref{fig:tr_cor} and \ref{fig:chromosphere} show the intensity, velocity and non-thermal velocity maps of the solar corona, transition region and chromosphere observed on 15 November 2017.
Based on the \ion{Fe}{xii} velocity map (Fig.~\ref{fig:tr_cor}), we defined the upflow region as the blue-shifted area with the Doppler velocity smaller than -5 km s$^{-1}$, and the active region core as the red-shifted area with the Doppler velocity larger than 5 km s$^{-1}$.

\begin{table*}[ht!]
\caption{Linear correlation coefficient between intensities of two close lines for the upflow region and active region core. The grey colour highlights the linear correlation coefficient lower than 0.5.}\label{tab:int_correlation}
\resizebox{\textwidth}{!}{\begin{tabular}{@{}lcccccc|ccccccc@{}}
\hline\hline
& \multicolumn{6}{c|}{Upflow region}                   &  & \multicolumn{6}{c}{Active region core}              \\ \hline
& 14 Nov & 15 Nov & 21 Nov & 22 Nov & 24 Nov & 25 Nov &  & 14 Nov & 15 Nov & 21 Nov & 22 Nov & 24 Nov & 25 Nov \\ \hline
\ion{Fe}{xiii}-\ion{Fe}{xiv}    & \cellcolor{green!0}{0.94}   & \cellcolor{green!0}{0.92}   & \cellcolor{green!0}{0.69}   & \cellcolor{green!0}{0.93}   & \cellcolor{green!0}{0.57}   & \cellcolor{green!0}{0.61}   &  & \cellcolor{green!0}{0.93}    & \cellcolor{green!0}{0.88}   & \cellcolor{green!0}{0.85}   & \cellcolor{green!0}{0.91}   & \cellcolor{green!0}{0.91}   & \cellcolor{green!0}{0.94}   \\
\ion{Fe}{xiii}-\ion{Fe}{xii}    & \cellcolor{green!0}{0.97}   & \cellcolor{green!0}{0.88}   & \cellcolor{green!0}{0.73}   & \cellcolor{green!0}{0.99}   & \cellcolor{green!0}{0.81}   & \cellcolor{green!0}{0.94}   &  & \cellcolor{green!0}{0.96}    & \cellcolor{green!0}{0.86}   & \cellcolor{green!0}{0.95}   & \cellcolor{green!0}{0.90}   & \cellcolor{green!0}{0.86}   & \cellcolor{green!0}{0.90}   \\
\ion{Fe}{xii} - \ion{Si}{iv}    & \cellcolor{green!0}{0.65}   & \cellcolor{green!0}{0.69}   & \cellcolor{green!0}{0.59}   & \cellcolor{green!0}{0.64}   & \cellcolor{green!0}{0.60}   & \cellcolor{green!0}{0.71}   &  & \cellcolor{green!0}{0.56}    & \cellcolor{green!0}{0.50}   & \cellcolor{green!0}{0.62}   & \cellcolor{green!0}{0.63}   & \cellcolor{gray!50}{0.49}   & \cellcolor{gray!50}{0.43}   \\
\ion{Si}{iv} - \ion{C}{ii}      & \cellcolor{green!0}{0.70}   & \cellcolor{green!0}{0.73}   & \cellcolor{green!0}{0.79}   & \cellcolor{green!0}{0.82}   & \cellcolor{green!0}{0.79}   & \cellcolor{green!0}{0.82}   &  & \cellcolor{green!0}{0.73}    & \cellcolor{green!0}{0.71}   & \cellcolor{green!0}{0.86}   & \cellcolor{green!0}{0.79}   & \cellcolor{green!0}{0.79}   & \cellcolor{green!0}{0.80}   \\
\ion{C}{ii} - Mg IIk3    & \cellcolor{green!0}{0.71}   & \cellcolor{green!0}{0.68}   & \cellcolor{green!0}{0.84}   & \cellcolor{green!0}{0.94}   & \cellcolor{green!0}{0.83}   & \cellcolor{green!0}{0.85}   &  & \cellcolor{green!0}{0.82}    & \cellcolor{green!0}{0.86}   & \cellcolor{green!0}{0.96}   & \cellcolor{green!0}{0.91}   & \cellcolor{green!0}{0.85}   & \cellcolor{green!0}{0.86}   \\
Mg IIk3 - Mg IIk2 & \cellcolor{green!0}{0.90}   & \cellcolor{green!0}{0.86}   & \cellcolor{green!0}{0.94}   & \cellcolor{green!0}{0.95}   & \cellcolor{green!0}{0.95}   & \cellcolor{green!0}{0.93}   &  & \cellcolor{green!0}{0.88}    & \cellcolor{green!0}{0.92}   & \cellcolor{green!0}{0.96}   & \cellcolor{green!0}{0.95}   & \cellcolor{green!0}{0.91}   & \cellcolor{green!0}{0.92}  \\ \hline

\end{tabular}}
\end{table*}

\begin{table*}[ht!]
\caption{Linear correlation coefficient between velocities of two close lines for the upflow region and active region core. Green color indicates the linear correlation coefficient higher than 0.5; light green  - correlation in range between 0.15 and 0.5; white  -correlation from -0.15 to 0.15 (lack of correlation); light red - negative correlation of -0.15 to -0.5.}\label{tab:vel_correlation}
\resizebox{\textwidth}{!}{\begin{tabular}{@{}lcccccc|ccccccc@{}}
\hline\hline
& \multicolumn{6}{c|}{Upflow region}                   &  & \multicolumn{6}{c}{Active region core}              \\ \hline
& 14 Nov & 15 Nov & 21 Nov & 22 Nov & 24 Nov & 25 Nov &  & 14 Nov & 15 Nov & 21 Nov & 22 Nov & 24 Nov & 25 Nov \\ \hline
\ion{Fe}{xiii}-\ion{Fe}{xiv}    & \cellcolor{green!15}{0.48}   & \cellcolor{green!100}{0.76}   & \cellcolor{green!15}{0.49}   & \cellcolor{green!100}{0.53}   & \cellcolor{green!15}{0.38}   & \cellcolor{green!15}{0.25}        && \cellcolor{green!100}{0.78}   & \cellcolor{green!100}{0.85}   & \cellcolor{green!15}{0.37}   & \cellcolor{green!100}{0.64}   & \cellcolor{green!100}{0.76}   & \cellcolor{green!15}{0.37}   \\
\ion{Fe}{xiii}-\ion{Fe}{xii}    & \cellcolor{green!100}{0.68}   & \cellcolor{green!100}{0.81}   & \cellcolor{green!100}{0.66}   & \cellcolor{green!100}{0.89}   & \cellcolor{green!100}{0.84}   & \cellcolor{green!100}{0.67}   && \cellcolor{green!100}{0.64}   & \cellcolor{green!100}{0.53}   & \cellcolor{green!100}{0.53}   & \cellcolor{green!100}{0.50}   & \cellcolor{green!100}{0.75}   & \cellcolor{green!100}{0.82}   \\
\ion{Fe}{xii} - \ion{Si}{iv}    & \cellcolor{red!0}{-0.05}    & \cellcolor{red!0}{-0.09}    & \cellcolor{green!0}{0.02}   & \cellcolor{green!0}{0.04}     & \cellcolor{red!25}{-0.30}    & \cellcolor{red!25}{-0.29}          && \cellcolor{green!15}{0.23}    & \cellcolor{green!15}{0.31}   & \cellcolor{green!15}{0.21}   & \cellcolor{green!15}{0.16}   & \cellcolor{green!0}{0.13}   & \cellcolor{green!15}{0.36}  \\
\ion{Si}{iv} - \ion{C}{ii}      & \cellcolor{green!15}{0.22}   & \cellcolor{green!100}{0.68}   & \cellcolor{green!100}{0.65}   & \cellcolor{green!100}{0.71}   & \cellcolor{green!100}{0.69}   & \cellcolor{green!15}{0.44}    && \cellcolor{green!15}{0.42}   & \cellcolor{green!100}{0.74}   & \cellcolor{green!100}{0.65}   & \cellcolor{green!15}{0.28}   & \cellcolor{green!15}{0.33}   & \cellcolor{green!100}{0.66}   \\
\ion{C}{ii} - Mg IIk3    & \cellcolor{green!0}{0.06}   & \cellcolor{green!15}{0.15}   & \cellcolor{green!15}{0.43}   & \cellcolor{green!15}{0.46}   & \cellcolor{green!15}{0.43}   & \cellcolor{green!0}{0.09}          && \cellcolor{green!15}{0.18}   & \cellcolor{green!15}{0.47}   & \cellcolor{green!15}{0.37}   & \cellcolor{green!15}{0.21}   & \cellcolor{green!15}{0.24}   & \cellcolor{green!15}{0.31}   \\
Mg IIk3 - Mg IIk2 & \cellcolor{green!100}{0.71}   & \cellcolor{green!100}{0.52}   & \cellcolor{green!100}{0.81}   & \cellcolor{green!100}{0.71}   & \cellcolor{green!100}{0.75}   & \cellcolor{green!100}{0.61}   && \cellcolor{green!100}{0.66}   & \cellcolor{green!100}{0.70}   & \cellcolor{green!15}{0.36}   & \cellcolor{green!100}{0.67}   & \cellcolor{green!100}{0.63}   & \cellcolor{green!100}{0.68}  \\ \hline

\end{tabular}}
\end{table*}

%Velocity
The velocity maps (Figs.~\ref{fig:tr_cor}) obtained from the coronal lines (\ion{Fe}{xii}, \ion{Fe}{xiii}, and \ion{Fe}{xiv}) are similar to each other.
These maps show two clear compact features - the blue-shifted upflow region and the red-shifted active region core.
On the contrary to the corona, the maps of the underlying layers (\ion{Si}{iv}, \ion{C}{ii}, \ion{Mg}{ii}) present a huge diversity of Doppler velocities and they are rich in numerous small-scale features.

%Intensity
The intensity maps (Figs.~\ref{fig:tr_cor},~\ref{fig:chromosphere}) show the bright active region core and significantly darker upflow region in the coronal lines.
In the transition region and chromosphere, the upflow region is only slightly darker than the active region core.
The \ion{Mg}{ii}~k2 line shows chromospheric network pattern (brighter than surroundings) and internetwork regions (darker than surroundings) that in the \ion{Mg}{ii} k2 velocity map corresponds to redshifted and blueshifted areas, respectively.

%Non-thermal velocity -how to describe these maps!
The non-thermal velocities (Figs.~\ref{fig:tr_cor}) were calculated only for optically thin lines observed in the corona (\ion{Fe}{xii}, \ion{Fe}{xiii}, and \ion{Fe}{xiv}) and the transition region (\ion{Si}{iv}). 
In the \ion{Fe}{xiii} and \ion{Fe}{xiv} lines, the non-thermal velocity is larger in the upflow region than in the active region core.
In the \ion{Fe}{xii} line, the non-thermal velocity in the whole active region core and one-third of the upflow region is stronger than 30 km s$^{-1}$.
In the rest of the upflow region, the non-thermal velocity is below 30 km s$^{-1}$.
%
%The non-uniform non-thermal velocity distribution in the upflow region suggests that there more one mechanisms are responsible for a non-thermal spectral line broadening.
The \ion{Fe}{xiii} and \ion{Fe}{xii} show stronger non-thermal velocity towards the boundary in the upflow region suggesting that there are more than one mechanism responsible for a non-thermal spectral line broadening.
The non-thermal velocity of \ion{Fe}{xii} 195.12\AA\ is generally larger in the AR core than in the upflow region.
In contrast, other corona lines (\ion{Fe}{xiii}, \ion{Fe}{xiv}) present the stronger average non-thermal velocity in the upflow region than in the active region core. 
It is very likely caused that the \ion{Fe}{xii} 195.12\AA\ line is blended with another line at 195.18\AA. It is density sensitive blend so the higher density active core implies the stronger line width than in the upflow region where density is lower.
Moreover, the upflow region and the active region core show similar non-thermal velocities in the transition region (\ion{Si}{iv}).

\subsection{Mutual relation between intensity in closest line}\label{sec:correlation_intensity}

%Calculation of the correlation coefficient
To quantify the relation between the plasma properties of pairs of spectral lines with close formation temperatures (e.g. \ion{Fe}{xii} and \ion{Fe}{xiii} lines), we used Pearson's linear correlation coefficient.
This coefficient has a value in the range from -1 (negative correlation) to 1 (positive correlation), while 0 defines the lack of correlation.
We calculated this correlation coefficient between the intensities of pairs of spectral lines with close formation temperatures separately for the upflow region (left side of Table~\ref{tab:int_correlation}) and the active region core (right side of Table~\ref{tab:int_correlation}).
These analyses showed a strong correlation for all ROIs in all observations. 
%%%%%%%%%%%%%%%%%%%%%%%

%The upflow region
In the upflow region, we found strong correlations (left side of Table~\ref{tab:int_correlation}), in a range from 0.57 to 0.99,  between intensities from the lines with close formation temperature.
In each observation, the relationship between the transition region and coronal lines (\ion{Si}{iv} - \ion{Fe}{xii}) is slightly weaker than that of other pairs, but the correlation coefficient is at least 0.57.

%The active region core
In the active region core, all pairs of spectral lines show the correlation coefficient around 0.5 or higher, but only in two cases, the correlation is below 0.5 (0.43, 0.49).
In general, we can distinguish two groups of the spectral line pairs with the higher correlation coefficient.
The coronal pairs present the correlation higher than 0.85.
The chromospheric and chromospheric-transition region pairs have correlation higher than 0.71.
Similarly to the upflow region, the correlation between the transition region and coronal lines are weaker than in other pairs; the correlation coefficient is below 0.5 in two observations.
The nearby limb observation from 24th and 25th November 2017 show a low correlation between \ion{Fe}{xii} and \ion{Si}{iv} lines (Table~\ref{tab:int_correlation}).
This low correlation can be a result of the strong projection effect. 
%

%Comparison -the upflow region and the active region core
The comparison of the intensity correlations of the pairs of lines with close formation temperature in the upflow region and active region core (Table~\ref{tab:int_correlation}) shows the same trends in both regions, for example two groups of spectral lines with the higher correlation (coronal and chromospheric-transition region) and the weaker correlation between the transition region and coronal line (\ion{Si}{iv} - Fe XII).
Moreover, we found a higher correlation for the upflow region lines than for the active region core.
The high correlation coefficients suggests a clear relation between the structures observed in the close formation temperatures and allows for analysis further plasma properties.

In general, we did not notice a significant relationship between the ROI positions on the solar disk and the correlation coefficient. Still, for large inclination, the projection effects can be important (e.g. observation from 14, 24, and 25 November 2017).

\subsection{Relationship between velocities in pairs of lines with close formation temperature}\label{sec:correlation_vel}

%General info
To quantify the dependence between the Doppler velocities in the closest formation temperature lines, we used the same correlation measure as in Sect.~\ref{sec:correlation_intensity}.
Table~\ref{tab:vel_correlation} presents the correlation between Doppler velocities for the upflow region (left side) and the active region core (right side).
%

%Upflow region -NEW
In the upflow region, the coronal (\ion{Fe}{xii}) and transition region (\ion{Si}{iv}) velocities show very weak correlation.
This is because, the \ion{Si}{iv} is mostly red-shifted, whereas the \ion{Fe}{xii} is blue-shifted.
%This is because, the \ion{Si}{iv} is red-shifted which indicates closed loops, whereas the \ion{Fe}{xii} is blue-shifted which indicates the open magnetic field.
%
We also found a weak correlation between the Doppler velocities of the \ion{C}{ii} and \ion{Mg}{ii} k3 lines.

The pair of the coronal lines (\ion{Fe}{xiii} and \ion{Fe}{xii}) are well correlated, but the correlation decreases towards higher temperatures (\ion{Fe}{xiv} and \ion{Fe}{xiii}).
We also noticed a strong correlation for the pair of the transition region lines (\ion{C}{ii}-\ion{Si}{iv}) and the pair of the chromospheric lines (\ion{Mg}{ii} k3 and \ion{C}{ii}).

%AR core
The velocities measured in the active region core and the upflow region presents almost the same trends. 
However, the correlation between the coronal (\ion{Fe}{xii}) and transition region (\ion{Si}{iv}) velocities is stronger in the active region core compare with the upflow region ones.

%Dependence correlation vs. position at the solar disk
Table~\ref{tab:vel_correlation} shows that the velocity correlations in the upflow region are stronger nearby the disk centre and weaker towards to limb.
We did not notice this effect in the active region core.

\subsection{Ratio between the average intensities of downflow and upflow regions.}\label{sec:ratio_intensity}
We calculated the ratio of the average active region core intensity to the upflow region intensity separately for each observation and at each spectral line. 
Table~\ref{tab:ratio_up_down} shows that the intensity ratio increases with the growing line formation temperatures and reaches its highest value in the corona.
In the corona, the active region core is always brighter than the upflow region (from 1.5 to 5.6 times) and usually at least two times brighter than in the transition region and chromosphere. 
The ratio sharply decreases to around one in the transition region; thus, there the active region core and the upflow region present similar brightness.
In the chromosphere, the upflow region usually is slightly brighter than the active region core, or both regions have similar brightness.
Thus, this similarity of intensity in chromosphere can suggest comparable chromospheric heating in the upflow region and the active region core, but further investigations are needed.

%=========TABLE=============

\begin{table}[ht!]
\caption{Ratio between the average intensities of the active region core and the upflow region.
The value are obtained for the six observations and six spectral lines covering the solar atmosphere from the corona throught the translation region to the chromosphere.
}\label{tab:ratio_up_down}
\resizebox{\columnwidth}{!}{\begin{tabular}{@{}lcccccc@{}}
\hline\hline
& 14 Nov&15 Nov&21 Nov&22 Nov&24 Nov&25 Nov \\ \hline
\ion{Fe}{xiv}  & \cellcolor{green!0}{4.39}  & \cellcolor{green!0}{5.74}   & \cellcolor{green!0}{3.54}   & \cellcolor{green!0}{1.61}   & \cellcolor{green!0}{4.47}   & \cellcolor{green!0}{4.99}   \\
\ion{Fe}{xiii} & \cellcolor{green!0}{2.92}   & \cellcolor{green!0}{2.58}   & \cellcolor{green!0}{2.29}   & \cellcolor{green!0}{1.40}   & \cellcolor{green!0}{2.31}   & \cellcolor{green!0}{2.71}   \\
\ion{Fe}{xii}  & \cellcolor{green!0}{2.96}   & \cellcolor{green!0}{2.58}   & \cellcolor{green!0}{2.41}   & \cellcolor{green!0}{1.48}   & \cellcolor{green!0}{2.19}   & \cellcolor{green!0}{2.48}   \\
\ion{Si}{iv}   & \cellcolor{green!0}{1.42}   & \cellcolor{green!0}{1.30}   & \cellcolor{green!0}{0.91}    & \cellcolor{green!0}{0.43}    & \cellcolor{green!0}{0.70}   & \cellcolor{green!0}{0.69}   \\
\ion{C}{ii}    & \cellcolor{green!0}{1.39}   & \cellcolor{green!0}{1.26}   & \cellcolor{green!0}{1.08}   & \cellcolor{green!0}{0.61}    & \cellcolor{green!0}{0.93}   & \cellcolor{green!0}{0.73}   \\
Mg IIk3 & \cellcolor{green!0}{1.20}    & \cellcolor{green!0}{1.25}   & \cellcolor{green!0}{1.18}   & \cellcolor{green!0}{0.51}    & \cellcolor{green!0}{0.98}   & \cellcolor{green!0}{0.76}   \\
Mg IIk2 & \cellcolor{green!0}{0.97}    & \cellcolor{green!0}{1.00}   & \cellcolor{green!0}{0.97}    & \cellcolor{green!0}{0.50}    & \cellcolor{green!0}{0.86}   & \cellcolor{green!0}{0.68} \\ \hline
\end{tabular}}
\end{table}

%==========================================================================
\begin{figure*}[ht!]
\centering
\includegraphics[scale=0.7]{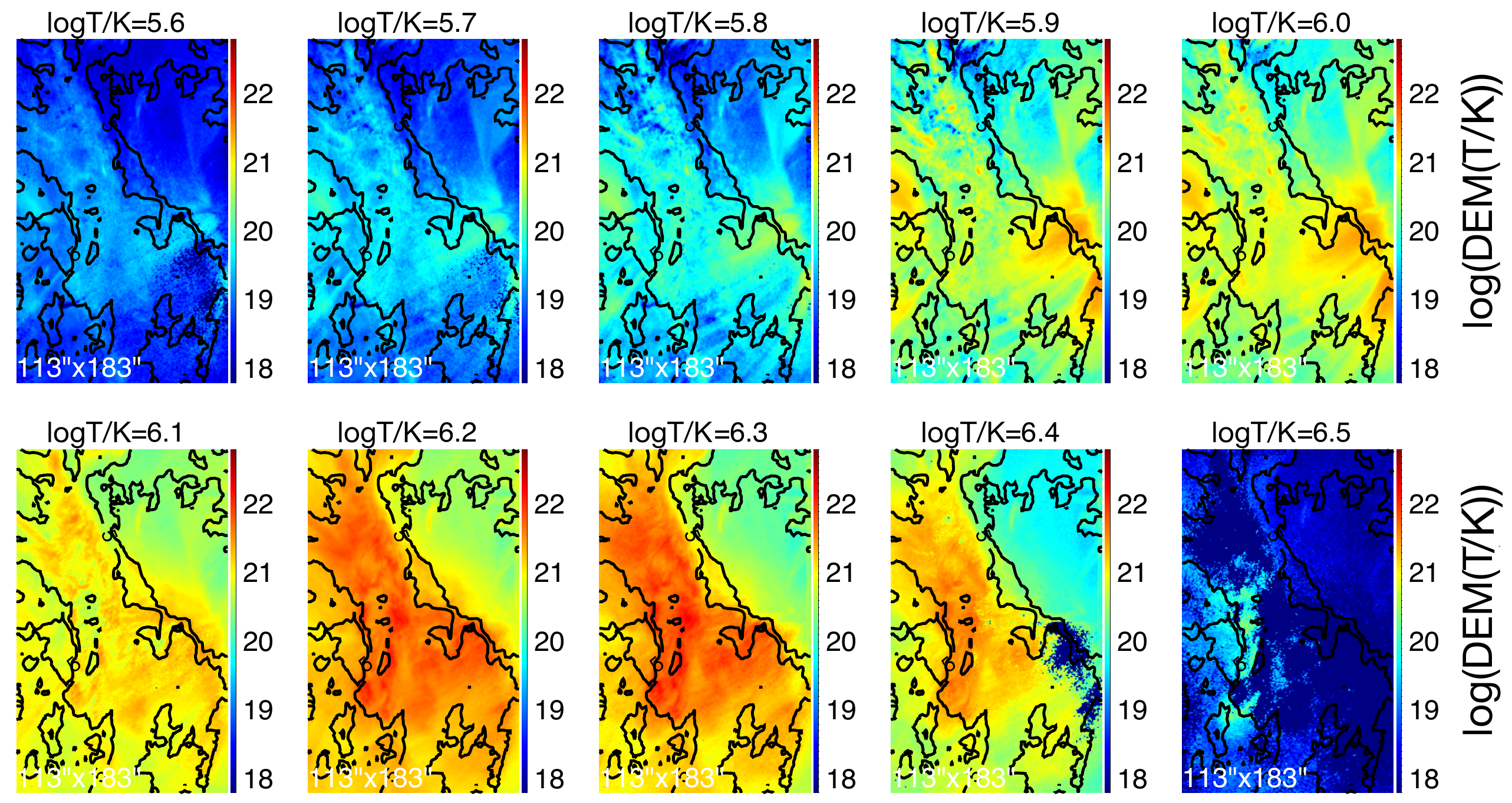}
\caption{
Thermal structure of the upflow region and the active region core for the observation from 15 November.
The panels display the differential emission measure (DEM) map at a temperature range from logT/K=5.6 to logT/K=6.5.
The field-of-view is the same as for IRIS raster data (Fig.~\ref{fig:tr_cor}).
The contours indicate the $\pm$5 km s$^{-1}$ plasma flow in the \ion{Fe}{xii} line indicating the upflow region and the active region core.}
\label{fig:dem}
\end{figure*}
%===========================================================================

%==========================================================================
\begin{figure}[ht!]
\centering
\includegraphics[scale=1.0]{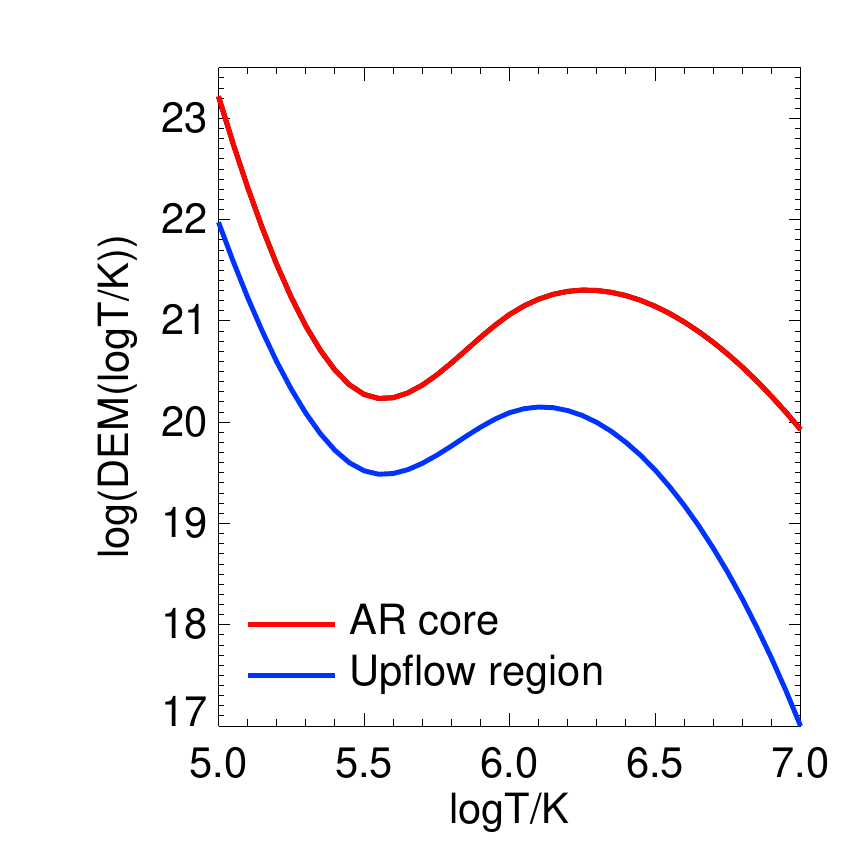}
\caption{
Differential emission measure (DEM) distribution for the active region core (red) and upflow region (blue) based on EIS observation of 12 spectral lines from 15 November. 
The lines show the DEM from the average intensity of the active region core and upflow region as defined in Sect.~\ref{sec:observation_maps} (Fig.~\ref{fig:tr_cor}).}
\label{fig:dem2}
\end{figure}
%===========================================================================

\subsection{Differential Emission Measure and density diagnostics}\label{sec:dem}
The Differential Emission Method (DEM) provides information about the plasma distribution in temperature throughout the atmosphere along the line-of-sight.
The DEM is defined:
$DEM=n^{2}_{e}\left( \frac{dT}{dh} \right)^{-1}$ and depends on $n_{e}$ -electron density, T -temperature, h-height along the line-of-sight. 
We performed a DEM analysis for the observation of 15 November based on SDO/AIA images of 94\AA, 131\AA, 171\AA, 193\AA, and 211\AA, 335\AA\, channels.
We used the regularized DEM inversion method prepared by \citet{Hannah2012}.
To this aim, we chose a set of near-simultaneous images obtained at 14:46:03 UT 15 November 2017.
The SDO/AIA images from all the above-mentioned channels were spatially aligned with single SDO/AIA pixel accuracy.
In our calculation, we took into account the photon noise and readout noise.

The DEM maps (Fig.~\ref{fig:dem}) show the strongest emission in the upflow region at logT/K=5.9-6.0 and in the active region core at logT/K=6.3.
This suggests that the upflow region is built from cooler plasma than the active region core.
Moreover, the upflow region presents less emission from the transition region temperatures than the active region core (logT/K=5.6-5.8). 
The emission distribution suggests that the dominant part of the upflow plasma is generated in the upper transition region or lower corona.
In the upflow region, the DEM increases towards the boundary with the active region core.

Additionally, we carried out a DEM analysis based on the observation of 15 November using Hinode/EIS data.
We used all available spectral lines:
\ion{He}{ii} (256.312\AA),  \ion{Fe}{viii} (185.213), \ion{Fe}{x} (184.536\AA, 257.262\AA), \ion{Fe}{xi} (180.401\AA), \ion{Fe}{xii} (186.880\AA,  193.509\AA, 195.119\AA), \ion{Fe}{xiii} (202.044\AA), \ion{Fe}{xiv} (274.203\AA), \ion{Fe}{xvi} (262.984\AA), \ion{Ca}{xvii} (192.858\AA).
\ion{He}{ii} line, optically thick, is used as a lower-limit constraint on the low temperature part of the DEM and it is not relevant for the coronal results we focus on.
We pre-processed Hinode/EIS data with a standard routine described in Sect.~\ref{tab:obs_overview}.
Based on the pre-processed data, the average intensity and their error were computed for the active region core and upflow region for 12 spectral lines, separately.
Then, we computed DEM(T) curves for the active region core and upflow region using the CHIANTI method \citep{DelZanna2021}.

The DEM curves shows the strongest emission in the upflow region at logT/K=6.1 and in the active region core at logT/K=6.25.
Moreover, the upflow region presents less emission than the active region core.
Therefore, the DEM analysis of EIS data are consistent with the DEM analysis based on SDO/AIA observation.
To clarify, the peak of the DEM for the active region core seems low (logT/K=6.3).
This is  because the intensities were averaged over the whole active core region so the contribution from higher temperature loops (logT/K=6.6 - see e.g. \citet{Warren2012}) is washed out.

To calculate the average density in the upflow region and active region core, we used the ratio of the \ion{Fe}{xii} lines (186.88\AA/195.119\AA) and SSW routine dens\_plot (\url{https://hesperia.gsfc.nasa.gov/ssw/packages/chianti/idl/ratios/ratio_plotter.pro}) based on CHIANTI \citep{DelZanna2021}.

We found a plasma density of $1.05\cdot10^9$cm$^{-3}$ in the upflow region.
In the active region core, the plasma density was $1.38\cdot10^{10}$cm$^{-3}$ (13 times larger than in the upflow region).

%00000000000000000000000000000000000000000000000000000000000

\section{Non-thermal velocity}\label{sec:nth_velocity}
We computed non-thermal velocity only for the optically thin lines: \ion{Si}{iv}, \ion{Fe}{xii}, \ion{Fe}{xiii} and \ion{Fe}{xiv}.

\subsection{Relation between Doppler velocity and non-thermal velocity}\label{sec:correlation_Doppler_nonth}

\begin{figure*}[ht!]
\centering
\includegraphics{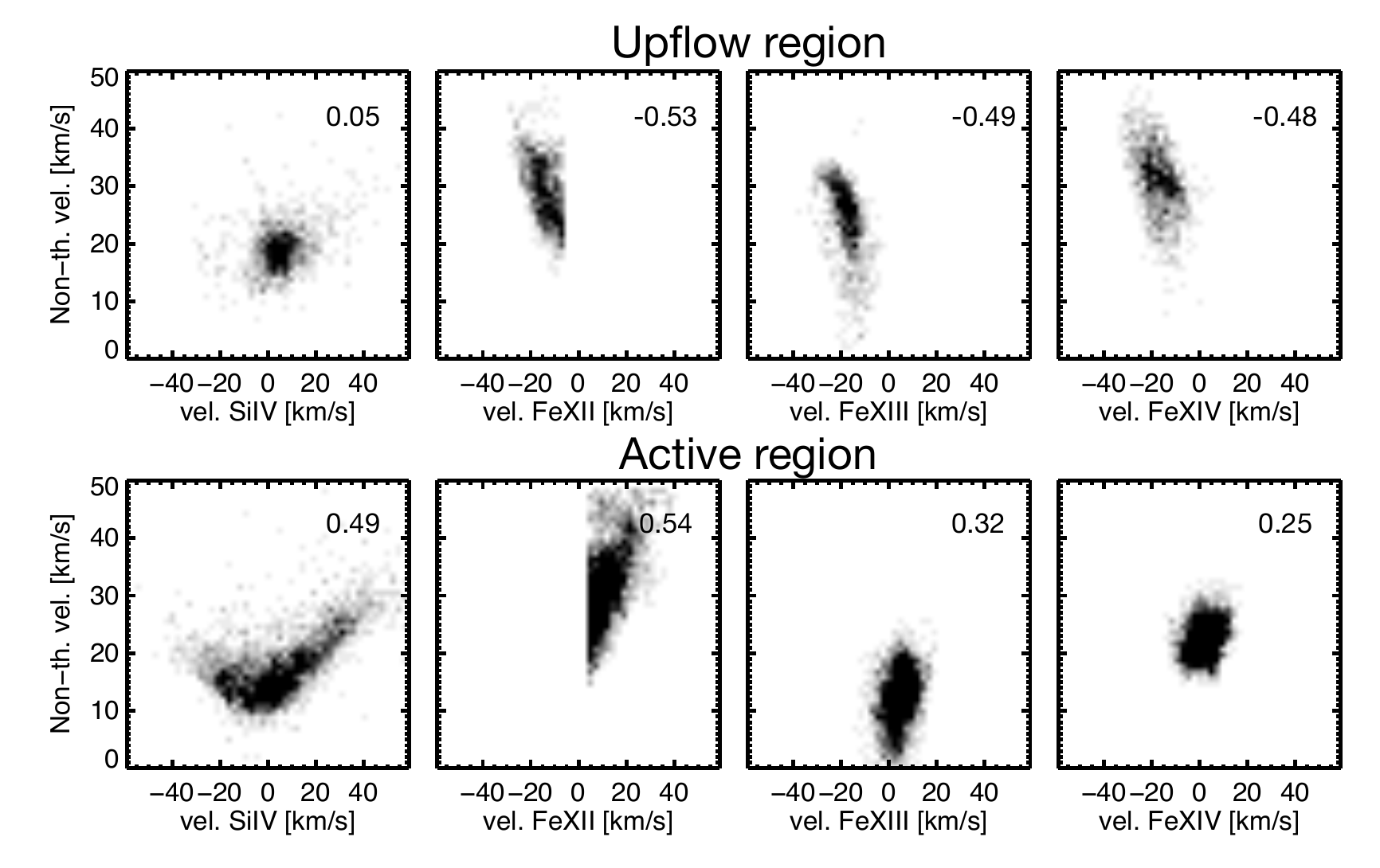}
\caption{
Probability density functions (PDFs) for the non-thermal velocity and Doppler velocity relation for the upflow region (upper panels) and the active region core (bottom panels) for the observation from 15 November 2017.
For the upflow region panels (top row), the number in the right corner shows the linear correlation coefficient.
For the active region core panels (bottom row), the number in the right corner presents a non-linear Spearman's rank correlation.
The panel of the \ion{Si}{iv} line in the active region core shows a dichotomy.
Thus, we additionally calculated the non-linear correlation coefficient separately for the negative, positive and absolute Doppler velocity and received -0.39, 0.74, and 0.62, respectively.
}
\label{fig:scatter_nonthvel_dvel}
\end{figure*}

To analyse the relationship between the Doppler velocity and non-thermal velocity, we prepared a 2D histogram of the probability density function (PDF) of \ion{Si}{iv}, \ion{Fe}{xii}, \ion{Fe}{xiii}, and \ion{Fe}{xiv} lines.
For the observation of 15 November, the PDF diagrams show a linear relationship between the non-thermal velocity and Doppler velocity for the \ion{Fe}{xii}, \ion{Fe}{xiii}, \ion{Fe}{xiv} lines in the upflow region.
We quantified this using Pearson's linear correlation coefficient.
For the active region core, the PDF diagrams show a non-linear correlation for the \ion{Si}{iv} line and the lack of correlation the coronal lines.
We used the non-linear Spearman's rank order correlation between for the active region core (Fig.~\ref{fig:scatter_nonthvel_dvel}).
The analogical analysis was carried out with all observations (Table~\ref{tab:dp_non_vel_correlation}).
In the upflow region (Fig.~\ref{fig:scatter_nonthvel_dvel} top, Table~\ref{tab:dp_non_vel_correlation} left side), the PDF diagrams show negative correlation in the coronal temperatures (\ion{Fe}{xii}, \ion{Fe}{xiii} and \ion{Fe}{xiv})  and lack of dependencies in the transition region (\ion{Si}{iv}). The negative correlation arises from the fact that we have chosen upflows to be represented by negative numbers, as compared with the positive numbers and correlation for downflows.
The strong negative correlation between the Doppler velocity and non-thermal velocity in the upflow region suggests that the processes responsible for non-thermal line broadening strongly influence upflow in the solar corona.
This correlation has been explained as a natural consequence of the superposition of high-speed upflow on the nearly static coronal background \citep{Tian2011}, or superposition of multiple upflows with slightly shifted velocities \citep{Doschek2007, Doschek2012}.
However, it is also possible in the other way that if there is an upflow, there may generated more turbulence on small scales.

In the active region core, we noticed the weak correlation or lack of correlation for \ion{Fe}{xii}, \ion{Fe}{xiii}, and \ion{Fe}{xiv} lines and positive correlation (of around 0.5) for \ion{Si}{iv} lines.
Thus, the closed magnetic field line topology of the active region core reduces the relation between the Doppler velocity and non-thermal velocity, possibly because the plasma motion is restricted within closed magnetic fields.
The open magnetic field lines topology of the upflow region allows for more freedom of the ionised plasma motion.
Therefore, we suggest that the non-thermal velocity is a result of the multiplicity of the line-of-sight velocities.
Moreover, stronger absolute Doppler velocity implies higher non-thermal velocity that can be related for example to increasing turbulence.

\subsection{Comparison of the average Doppler and non-thermal velocities in the upflow region and the active region core}\label{sec:average_velocities}

\begin{figure*}[ht!]
\centering
\includegraphics{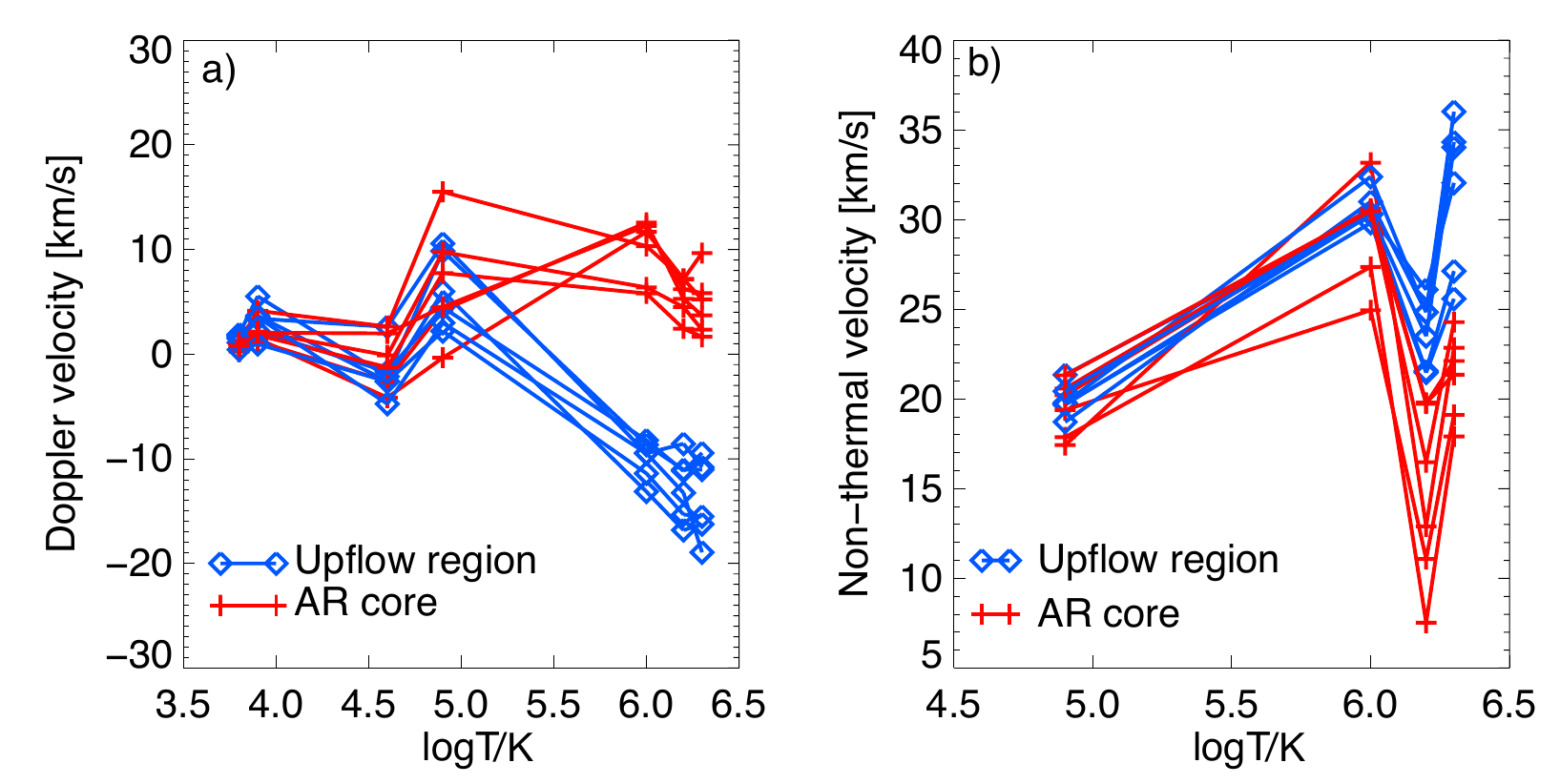}
\caption{
The average Doppler velocities (panel a) and the non-thermal velocities (panel b) ordering with the growing line formation temperatures.
The panels shows velocities calculated in the upflow region (line blue) and the active region core (red line) for six observations.}
\label{fig:mean_velocities}
\end{figure*}

We compared the average Doppler and non-thermal velocities computed separately for the upflow region and the active region core.
We ordered the average Doppler velocity according to the increasing line formation temperature and presented in Fig.~\ref{fig:mean_velocities}a for the upflow region (blue) and the active region core (red).
In the solar corona, the upflow region is always blueshifted, and the active region core is redshifted.
We found almost the same velocities for the upflow region and the active region core in the transition region and chromosphere.
The average Doppler velocities distribution in the upflow region suggests that at least one upflow mechanism is located between \ion{Si}{iv} (upper transition region) and \ion{Fe}{xii} (lower corona) lines formation temperature.
Several previous studies \citep{Teriaca1999, Peter1999, Dadashi2011} present the distribution of the average Doppler velocity with line formation temperature for active regions, quiet Sun, or the whole solar disk.
However, only few papers focus on the average Doppler velocity distribution in an upflow region  \citep[e.g.][]{DelZanna2008,Polito2020}.

%Non-thermal velocity
We analysed non-thermal velocities in the same way as the Doppler velocities.
They are presented in Fig.~\ref{fig:mean_velocities}b and ordered with the growing line formation temperatures (Fig.~\ref{fig:mean_velocities}b).
In the coronal lines, the non-thermal velocities are higher in the upflow region than in the active region core.
Moreover, in the coronal lines, the non-thermal velocities show a huge diversity from less than 10 km s$^{-1}$ in the active region core to almost 40 km s$^{-1}$ in the upflow region.
There is a general resemblance with Doppler velocity, the non-thermal velocity for \ion{Fe}{xiii} is usually lower than in \ion{Fe}{xii} and \ion{Fe}{xiv} lines.
In the transition region (\ion{Si}{iv} line), the non-thermal velocities are in a range of 17-22 km s$^{-1}$ for both the upflow region and the active region core.
The analysis of the average non-thermal velocity suggests that the magnetic field topology has a strong influence on the non-thermal broadening.
In the solar corona, the non-thermal velocity is stronger in the upflow region dominated by the open magnetic field lines, compared to the active region core governed by closed corona loops.
In general, the relation between the non-thermal velocity and the line formation temperature was investigated in several papers \citep{Chae1998, Peter2001, Brooks2016}.
However, we did not find a distribution of the non-thermal velocity with the line formation temperature for an upflow region in previous papers.

\begin{table*}[h]
\caption{Linear correlation coefficient between Doppler and non-thermal velocities for the upflow region and active region core. Green color indicates the linear correlation coefficient higher than 0.5; light green  - correlation in range between 0.15 and 0.5; white  -correlation from -0.15 to 0.15 (lack of correlation); light red - negative correlation of -0.15 to -0.5; red -strong negative correlation between -0.5 and -1; light blue -Spearman non-linear correlation coefficient between 0.15 to 0.5 and blue -Spearman non-linear correlation coefficient between 0.5 and 1.0.
}\label{tab:dp_non_vel_correlation}
\resizebox{\textwidth}{!}{\begin{tabular}{@{}lcccccc|ccccccc@{}}
\hline\hline
& \multicolumn{6}{c|}{Upflow region}                   &  & \multicolumn{6}{c}{Active region core}              \\ \hline
& 14 Nov & 15 Nov & 21 Nov & 22 Nov & 24 Nov & 25 Nov &  & 14 Nov & 15 Nov & 21 Nov & 22 Nov & 24 Nov & 25 Nov \\ \hline
\ion{Fe}{xiv}    & \cellcolor{red!15}{-0.26}   & \cellcolor{red!15}{-0.48}   & \cellcolor{red!100}{-0.73}   & \cellcolor{red!100}{-0.76}   & \cellcolor{red!100}{-0.56}   & \cellcolor{red!15}{-0.33}         &  & \cellcolor{red!0}{-0.04}     & \cellcolor{cyan!15}{0.25}   & \cellcolor{cyan!0}{0.03}   & \cellcolor{cyan!15}{0.22}   & \cellcolor{cyan!0}{0.07}   & \cellcolor{red!0}{-0.06}   \\
\ion{Fe}{xiii}   & \cellcolor{red!15}{-0.44}   & \cellcolor{red!15}{-0.49}   & \cellcolor{red!15}{-0.30}    & \cellcolor{red!100}{-0.68}   & \cellcolor{red!100}{-0.50}   & \cellcolor{red!0}{-0.16}          &  & \cellcolor{cyan!15}{0.27} & \cellcolor{cyan!15}{0.32}   & \cellcolor{cyan!15}{0.35}   & \cellcolor{cyan!100}{0.58}   & \cellcolor{cyan!15}{0.42}   & \cellcolor{cyan!15}{0.40}   \\
\ion{Fe}{xii}    & \cellcolor{red!100}{-0.69}   & \cellcolor{red!100}{-0.53}   & \cellcolor{red!15}{-0.45}   & \cellcolor{red!100}{-0.60}     & \cellcolor{red!100}{-0.79}    & \cellcolor{red!100}{-0.50}    &  & \cellcolor{cyan!0}{0.11} & \cellcolor{cyan!100}{0.54}   & \cellcolor{cyan!0}{0.04}   & \cellcolor{cyan!15}{0.24}   & \cellcolor{cyan!0}{0.08}   & \cellcolor{cyan!100}{0.60}  \\
\ion{Si}{iv}     & \cellcolor{red!0}{-0.03}    & \cellcolor{green!0}{0.05}   & \cellcolor{green!15}{0.37}  & \cellcolor{green!15}{0.44}   & \cellcolor{green!15}{0.16}   & \cellcolor{red!0}{-0.04}           &  & \cellcolor{cyan!15}{0.15}  & \cellcolor{cyan!15}{0.49}   & \cellcolor{cyan!100}{0.60}   & \cellcolor{cyan!100}{0.64}   & \cellcolor{cyan!100}{0.62}   & \cellcolor{cyan!15}{0.31}   \\ \hline
\end{tabular}}
\end{table*}

%======================================
%======================================

\section{Spectra classification}\label{sec:spectra_class}

We used an unsupervised algorithm known as the k-means clustering algorithm to identify groups of similar spectral profiles \citep{macqueen1967}.
For a detailed description of an application on IRIS spectra, see \citet{Panos2018}.
The IRIS level-2 raster data were binned to the Hinode spatial scale, but the spectral resolution was not modified.
Each spectrum was then normalised by its maximum intensity value before using the k-means algorithm separately on \ion{Mg}{ii}, \ion{C}{ii}, and \ion{Si}{iv} spectra.
The number of groups is often an ill defined parameter, however, there exists several methods that can provide an estimation of this quantity.
In this regard, we use the elbow method, which tracks the  decrease in variance while incrementally increasing the group number \citep{Thorndike1953}.
We obtained 8 groups for \ion{Mg}{ii}, 8 groups for \ion{C}{ii} and 8 groups for \ion{Si}{iv}, with each group containing spectra with a similar shape.

\begin{figure*}[ht!]
\centering
\includegraphics[scale=0.6]{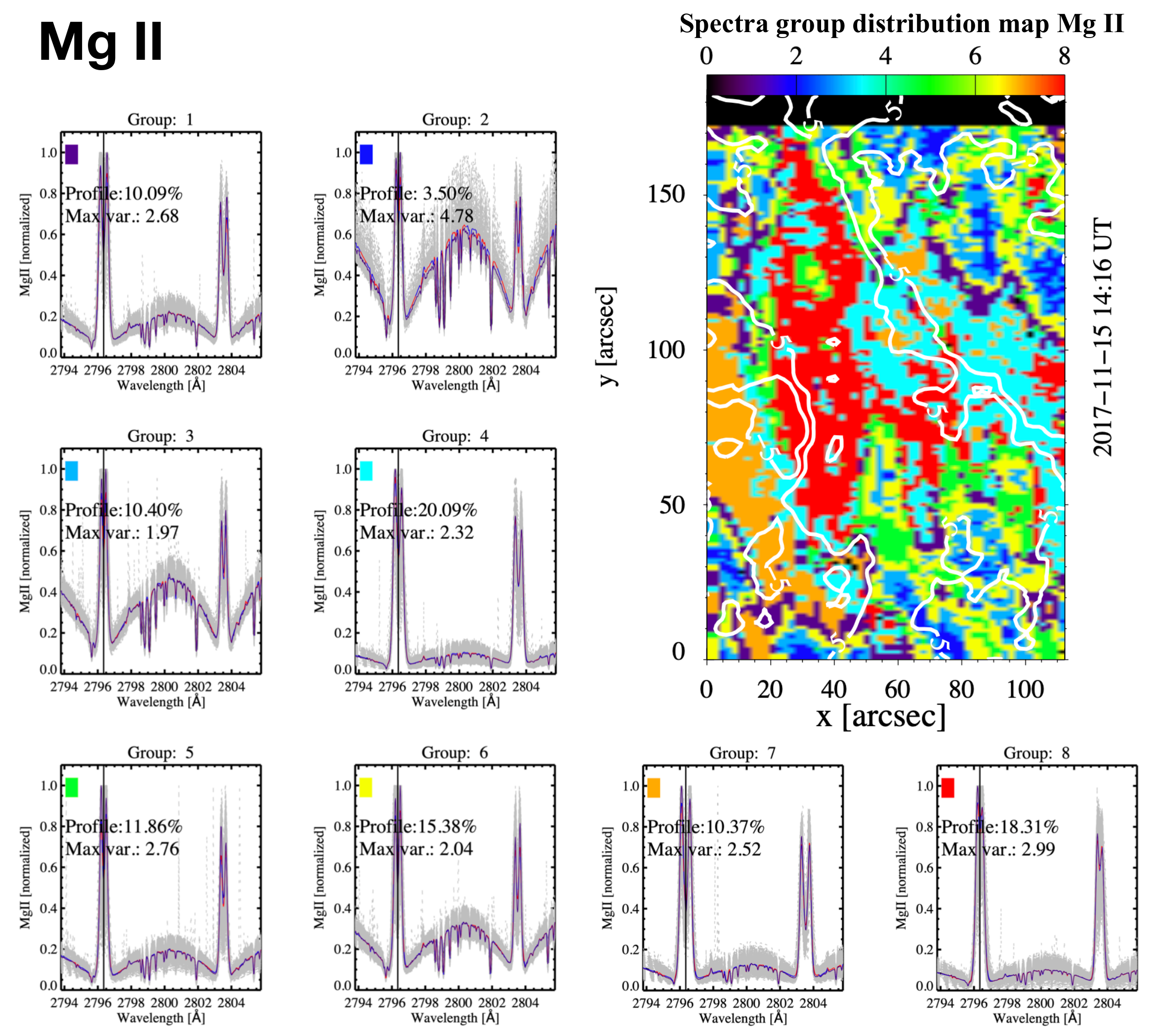}
\caption{
Classification of all \ion{Mg}{ii} spectra in the FOV into 8 groups using the k-means algorithm.
The small panels show the most characteristics spectra classified into 8 groups.
The blue line indicates the centroid of k-means, the blue-red line presents the average spectral line profile, and the grey lines show the variance of all profiles in the group.
The color of the rectangle in the spectra corresponds to the color indicating the location of the spectrum in the distribution map (top right panel).
The white contours of the distribution map outline the $\pm$5 km s$^{-1}$ plasma flow in \ion{Fe}{xii}, thus indicating the position of the upflow region and the active region core.}
\label{fig:sp_class}
\end{figure*}

Figure~\ref{fig:sp_class} shows the representative \ion{Mg}{ii} profiles of the 8 groups and the map presents the distribution of the groups in the active region.
Group 1-4 (violet-light blue) are characteristics for the upflow region; among them group 4 (light blue) is the most common.
Group 8 (red) is the most representative of the active region.
In groups 1-3, the right peaks of \ion{Mg}{ii} k2 and h2 are stronger than the left one, which indicates the blueshift, but in the spectra of groups 4,5 and 8, the left peak is stronger, which indicates redshift \citep{Pereira2013}.
Therefore, in the chromosphere of the upflow region we find blueshifted and redshifted areas.
Moreover, in the upflow region, the position of group 4 (redshifted) corresponds well to the area with a non-thermal velocity of the \ion{Fe}{xii} line stronger than its surroundings (Fig.~\ref{fig:tr_cor}).
Therefore, we suggest that group 4 points a region with the reconnection between active region closed loops and open magnetic field lines of the upflow region, especially that this group is located on the border between the active region core and the upflow region.

Comparing groups 4 and 8, we found that the upflow region spectra present deeper central reversal at the \ion{Mg}{ii} k3 and \ion{Mg}{ii} h3 lines centre than the spectra in the active region core.
However, further investigations are needed to explain the central reversal difference between the upflow region and active region core.

The maps of the groups of the characteristic spectra distribution for \ion{C}{ii} and \ion{Si}{iv} lines are discussed in more depth in the Appendix~\ref{app:app1}.
These maps suggest that the spectra groups of \ion{Mg}{ii} lines can correspond with the location and shape of the spectra groups of \ion{C}{ii} lines and \ion{Si}{iv} lines.
This  implies that the structures and physical processes in the chromosphere and transition region are dependent on each other.
However, the investigation of other regions are necessary to test the relation between the spectral groups from the different spectral lines.
%

%==========================================================================
%===========================================================================

\section{Discussion and conclusion}\label{sec:discussion}

In this paper, we have analysed the plasma properties in the upflow region and the active region core.
The plasma properties in the upflow region have been investigated before by many authors (Sect.~\ref{sec:intro}).
We extended the analysis to a broader temperature range covering the solar atmosphere from the chromosphere to the corona.
We summarise the most important results of our analysis below:
\begin{itemize}
\item The Doppler velocities between adjacent temperature ions in the upflow region show no correlation between the corona and the transition region but a very strong correlation in the corona. The upflow plasma is dominant in the corona.
\item The ratio between intensity of the core and upflow region is high in the upper corona reaching over 5, whereas it is around 1 in the chromosphere. The upflow region has low corona density, but the density is similar between the core and upflow region in the lower atmosphere.
\item The DEM analysis indicates the upflow region has a dominant temperature of log T=5.9-6.1, whereas the active region core is dominant from log T=6.2-6.4. The upflow regions are cool, and the emission is strongest closest to the boundary with the active region.
\item The Doppler velocity and non-thermal velocity show a strong correlation in the coronal lines in the upflow region – a correlation that is not seen in the active region core. The strength of this relationship may be due to the open field lines in the upflow region. 
\item The relationship between both Doppler velocity and non-thermal velocity with temperature has been explored before (e.g. \citet{Brooks2016}). In our case the upflow region deviates in its temperature behaviour above 1MK. 
\item In the upflow region of the chromosphere, there are both red and blueshifted regions, with the redshifted region being more dominant closer to the active region boundary. This is where the strongest non-thermal velocity is seen in the corona. 
\item In the upflow region, for each of the investigated spectral lines (\ion{Mg}{ii}, \ion{C}{ii} and \ion{Si}{iv}) we found several groups of the spectral lines that spatially correspond to each other.
Thus, we suggest that more than one process is responsible for the upflow.
\end{itemize}

\begin{figure*}[ht!]
\centering
\includegraphics[scale=0.4]{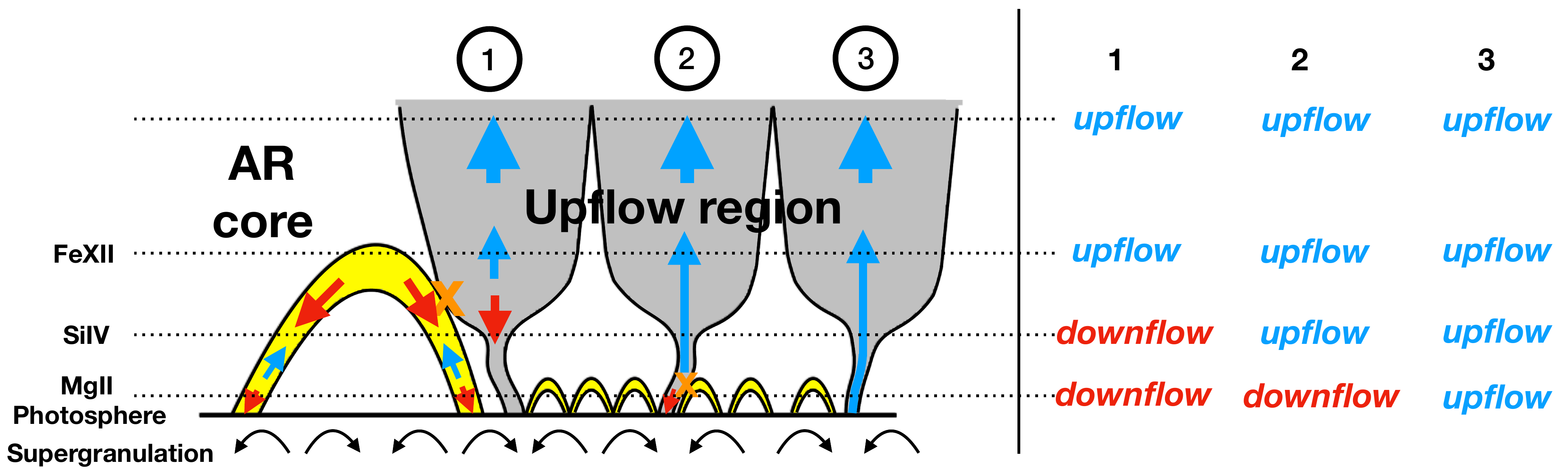}
\caption{
Mechanisms of the plasma upflow formation in the solar atmosphere. The sketch shows the magnetic field topology of the active region core and the upflow region (left side) and the description of the plasma flow in each region that corresponds to the blue (upflow) and red (downflow) arrows on the left side. The active region core is built with the hot coronal loop (yellow). The open magnetic field topology is characteristic for the upflow region (grey). In the chromosphere exists numerous small-scale loops (yellow). The numbers describe the processes responsible for the upflow generation. (1) The reconnection between the closed loop and open magnetic field lines (orange X) in the upper transition region or lower corona. Due to reconnection, plasma is injected towards the corona and chromosphere. The magnetic field increases towards the chromosphere and can decelerate the plasma flow in this direction. (2) The reconnection between the small-scale chromospheric loop and open magnetic field lines (orange X). Due to reconnection, plasma is injected towards the lower chromosphere and the transition region. The expanding open magnetic fields lines allow the plasma propagate further to the corona. (3) The open magnetic field structures allows plasma to escape from the chromosphere through the transition region to the solar corona. This process is accelerated by waves propagated in the lower solar atmosphere.}
\label{fig:sketch}
\end{figure*}

Based on the previous papers (Sect.~\ref{sec:intro}) and our results, we suggest three parallel processes that together can generate the plasma upflow (Fig.~\ref{fig:sketch}):
\begin{enumerate}
\item the interchange reconnection between the open magnetic field lines of the upflow region and closed magnetic field lines (loops) of the active region core in the solar corona;
\item the reconnection between the small-scale loops in the chromosphere and the open magnetic field lines;
\item expansion of the open magnetic field lines from the photosphere to the corona allows for the plasma upflow due to the waves in the lower solar atmosphere.
\end{enumerate}

We review each of these three options and describe in which location and why our observations are consistent with each scenario.
Previous papers \citep[e.g.][]{Peter2001, Tian2009} present cartoons of the coronal funnels in the quiet Sun or coronal holes.
Our sketch (Fig.~\ref{fig:sketch}) extends these cartoons for an upflow region considering different upflow scenarios.

\subsection{Interchange reconnection between the loops and the open magnetic field lines in the solar corona}

%Observation 
In the upflow region, our analysis shows strong blueshift in the corona and redshift in the transition region (Sect.~\ref{sec:observation_maps}).
This implies no correlation between the coronal and transition region velocities in the upflow region (Sect.~\ref{sec:correlation_vel}) as noticed in the previous studies for example \citet{Warren2011, UgarteUrra2011}.

% DEM, non-thermal velocity, and redshift in the chromosphere
The DEM analysis (Sect.~\ref{sec:dem}) indicates the stronger emission closest to the boundary between the upflow region and the active region.
This is where the strongest non-thermal velocity is seen in the corona (Sect.~\ref{sec:observation_maps}), and the redshifted region dominates in the chromosphere.

%Mechanism description
We suggest that reconnection occurs between the open magnetic field lines and closed magnetic field loops in the upper transition region or the lower solar corona (Fig.~\ref{fig:sketch}, scenario 1).
As a result of reconnection, plasma is injected towards the solar corona and photosphere that generates upflow (blueshift) in the corona and downflow (redshift) in the transition region, respectively.
The chromospheric redshift can be interpreted as the footprint of the plasma downflow generated during the reconnection.
However, the increasing magnetic field concentrations towards the photosphere decelerate this plasma; thus, the chromospheric line's plasma velocities are closer to 0 km/s than in the transition region.
The interchange reconnection should be especially efficient close to the border between the active region core and the upflow region where the open and closed magnetic field lines meet each other.
As a result of the reconnection, the emission increases, hence the DEM increases too. 
The turbulence related to reconnection can increase non-thermal line broadening \citep{Gordovskyy2016}.

We received the plasma speed in the upflow region that is significantly lower than the Alfv\'en speed \citep{Regnier2008}.
However, measured reconnection upflow speed, even during flares, are not at the Alfv\'en speed too that is confirmed in various papers for example \citet{Wang2017}.
Moreover, the determining of the Alfv\'en speed is tricky as well \citet{Regnier2008}.
Chromospheric evaporation also shows the speed lower than the Alfv\'en speeds as presented by \citet{Warren2006} used a multi-thread model.
This discrepancy can be explained that within spectral line broadening can exist multiple different blue shift components.

%\citep{Harra2008, Doschek2008, Bryans2010, Doschek2010, Doschek2012, Doschek2012}

\subsection{The reconnection between the small-scale loops and the open magnetic field lines in the chromosphere}
%Observation
The Doppler velocity maps of the transition region lines in the upflow region are dominated by the redshift structures, but these maps consists also small-scale blueshifted patches (Sect.~\ref{sec:observation_maps}).
The blueshifted structures in the transition lines of the upflow region were noticed in the previous papers, for example \citet{DelZanna2008, He2010}.
However, some of the transition region blueshifted patches spatially correspond to the redshifted area in the chromospheric lines.
Moreover, our analysis showed a strong correlation of the Doppler velocities of \ion{Si}{iv} - \ion{C}{ii} lines and line, a weaker correlation for \ion{C}{ii} - \ion{Mg}{ii} k3 lines and stronger for \ion{Mg}{ii} k3 - \ion{Mg}{ii} k2 lines (Sect.~\ref{sec:correlation_vel}).

%Mechanism description
We suggest, the reconnection between the chromospheric loops and the open magnetic field lines (Fig.~\ref{fig:sketch}, scenario 2) causes the upflow above the reconnection place (above the upper chromosphere) and the downflow below it (middle chromosphere and below).

\subsection{Expansion of the open magnetic field lines from the chromosphere to the corona}
%Observation
The Doppler velocity maps show spatially corresponding small-scale blueshifted patches in the chromosphere (\ion{Mg}{ii}) and transition region (\ion{Si}{iv}) of the upflow region.
These blueshifted areas are especially well visible in the internetwork of \ion{Mg}{ii} k2 (Sect.~\ref{sec:observation_maps}).

%Mechanism description
We suggest that in the upflow region, the open magnetic field expands from the photosphere to the corona. 
Therefore, plasma can easily escape from the chromosphere to the corona and further along the open magnetic field lines even when there is no reconnection.
The propagation of the waves in the lower solar atmosphere can accelerate the plasma upflow (Fig.~\ref{fig:sketch}, scenario 3).
Thus, a blueshift observed from the chromosphere via transition region to the solar corona is the most characteristic pattern of the upflow generated by the expanding open flux tubes mechanism.

%Jet-spicule explanation
\citet{Polito2020} found significant differences in spectral signatures below the upflows compared to the core of an AR observed by IRIS. For example, they noticed blueshifts in the chromosphere (\ion{Mg}{ii}, \ion{C}{ii}), and smaller redshifts and the presence of blueshifts in the transition region (\ion{Si}{iv}) below the coronal upflows (\ion{Fe}{xii}).
This connection between the chromosphere, transition region, and corona suggests that the expanding magnetic field scenario can allow jets and spicules to create the upflow.

\subsection{Comparison of the upflow region and the active region core}

The spectral profile shape classification (Sect.~\ref{sec:spectra_class}) shows a clear difference between the upflow region and the active region core in the chromosphere (\ion{Mg}{ii}, \ion{C}{ii}) and transition region (\ion{Si}{iv}).
We distinguished several groups of the \ion{Mg}{ii} profiles.
Profiles within the same group are similar, but profiles from different groups consists of significantly different spectra.
We noticed the stronger central reversal of \ion{Mg}{ii} k3 and \ion{Mg}{ii}h3 in the upflow region than in the active region core.
To understand the \ion{Mg}{ii} line properties and also explain the deep central reversal differences between the upflow region and the active region core, we require additional observations and simulations.

%Intensity ratio +DEM -> the upflow region -low dense, cooler
The ratio between the intensity of the core and upflow region (Sect.~\ref{sec:ratio_intensity}) shows that the coronal plasma density is lower in the upflow region than in the active region core.
Moreover, the DEM analysis (Sect.~\ref{sec:dem}) indicates that the coronal plasma is cooler in the upflow region than in the active region core.
The open magnetic field structures dominate the upflow region.
But the active region core is rich in the coronal loops, closed magnetic field structures preventing the plasma from escaping.
 Therefore, in the upflow region the plasma can escape more readily into space, hence, the density and temperature of the coronal plasma in the upflow region are lower than in the active region core.
However, in the chromosphere and the transition region, the feeds of the different coronal structures look the same \citep{Peter2001}, hence the average intensity of the active region core and upflow region are similar.

The Doppler and non-thermal velocity change with the growing line formation temperature (Sect.~\ref{sec:average_velocities}). 
The increase of the negative Doppler velocity in the upflow region with temperature is due to the plasma acceleration in the open magnetic field lines.
Considering the two-component \citep{Tian2011} or three-component scenarios \citep{McIntosh2012}, there is an alternative interpretation of this observational result: the different relative contributions of the fast upflow component and the cooling downflow component at different temperatures \citep{Wang2013}.
The deviation of the non-thermal velocity in the temperature above 1MK suggests that different mechanisms can be responsible for the non-thermal heating in the upflow region and the active region core.

The strong correlation between the Doppler and non-thermal velocity (Sect.~\ref{sec:correlation_Doppler_nonth}) observed in the coronal line of the upflow region and a lack of this correlation in the active region core suggest that the magnetic field topology can have a significant influence on the non-thermal heating mechanism.
It is also possible that in these two region, there are different non-thermal heating mechanisms at work.
However, it is difficult to point out which non-thermal heating mechanism plays a dominant role in our ROIs.
Thus, further investigation of the non-thermal velocity is necessary.

%===========================================================================

%=============================================================================================================================================
%=============================================================================================================================================
%=============================================================================================================================================
%=============================================================================================================================================

% \bibliographystyle{aa} % style aa.bst
 % your references Yourfile.bib

\begin{acknowledgements}
   This work is supported by Swiss National Science Foundation - SNF. The work of DHB was performed under contract to the Naval Research Laboratory and was funded by the NASA Hinode program. SDO data  are  courtesy of NASA/SDO and the AIA, EVE, and HMI science teams. IRIS is a NASA small explorer mission developed and operated by LMSAL with mission operations executed at NASA Ames Research Center and major contributions to downlink communications funded by ESA and the Norwegian Space Centre. Hinode is a Japanese mission developed and launched by ISAS/JAXA, with NAOJ as domestic partner and NASA and STFC (UK) as international partners. It is operated by these agencies in co-operation with ESA and NSC (Norway). 
\end{acknowledgements}

\begin{appendix}
\section{Spectra classification}\label{app:app1}
\begin{figure*}[ht!]
\centering
\includegraphics[scale=0.73]{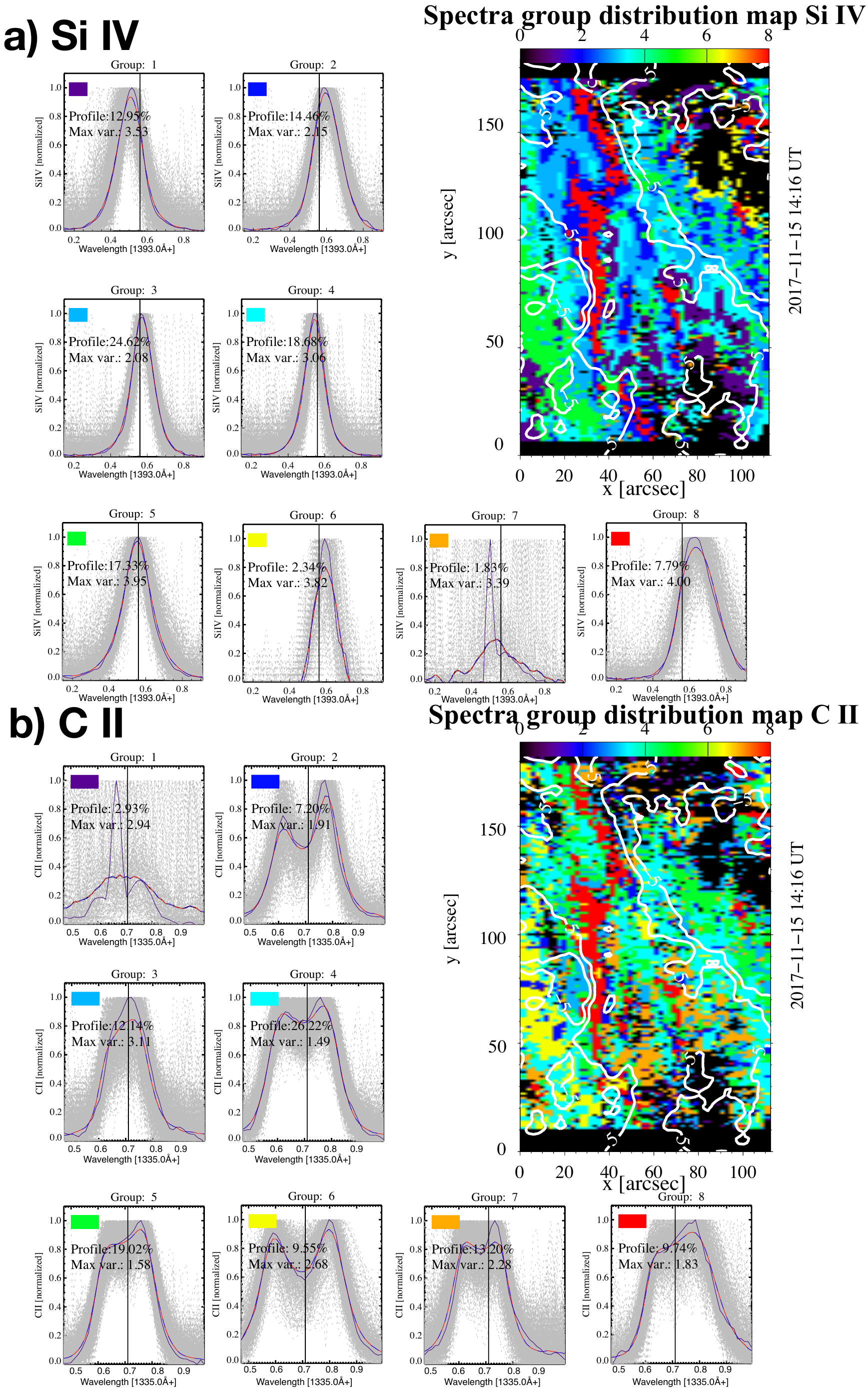}
\caption{
The groups of the most characteristics spectra of \ion{Si}{iv}(a) and \ion{C}{ii} (b).
The small panels show the most characteristics spectra in 8 groups for \ion{Si}{iv} (a) and 8 groups of \ion{C}{ii} (b).
The blue line indicates the most characteristic profile (the centroid of k-means), the blue-red line presents the average spectral line profile, and the grey line presents all profiles in the group.
The colour of the rectangular box (right top of the spectra panel) indicates the spectrum's location in the distribution map (top right panel).
The white contours of the distribution map present the $\pm$5 km s$^{-1}$ plasma flow in \ion{Fe}{xii}, thus indicating the upflow region's position and the active region core.}
\label{fig:spectra_c2si4}
\end{figure*}

We used the k-means method to classify the groups of similar spectral lines in an analogous way as in Sect.~\ref{sec:data_analysis}.
Figure~\ref{fig:spectra_c2si4} shows the representative spectra and their distribution map for \ion{Si}{iv} (a) and \ion{C}{ii} (b) lines.
These maps show that the upflow region group of spectral lines in Fig.~\ref{fig:sp_class} (group 4, light blue) clearly corresponds to groups 3 and 4 (blue) in \ion{Si}{iv} (Fig.~\ref{fig:spectra_c2si4}a) and groups 4 (blue) and 5 (green)  in \ion{C}{ii} (Fig.~\ref{fig:spectra_c2si4}b).
The relationship between the active region core groups is even stronger, group 8 (red) in Fig.~\ref{fig:sp_class} corresponds to \ion{Si}{iv} group 8 (red) in Fig.~\ref{fig:spectra_c2si4}a and \ion{C}{ii} group 8 (red) in Fig.~\ref{fig:spectra_c2si4}b.

%==========================================================================
%===========================================================================

\end{appendix}


\begin{thebibliography}{69}
\expandafter\ifx\csname natexlab\endcsname\relax\def\natexlab#1{#1}\fi

\bibitem[{{Abbo} {et~al.}(2016){Abbo}, {Ofman}, {Antiochos}, {Hansteen},
  {Harra}, {Ko}, {Lapenta}, {Li}, {Riley}, {Strachan}, {von Steiger}, \&
  {Wang}}]{Abbo2016}
{Abbo}, L., {Ofman}, L., {Antiochos}, S.~K., {et~al.} 2016, \ssr, 201, 55

\bibitem[{{Athay} \& {Dere}(1991)}]{Athay1991}
{Athay}, R.~G. \& {Dere}, K.~P. 1991, \apj, 381, 323

\bibitem[{{Baker} {et~al.}(2017){Baker}, {Janvier}, {D{\'e}moulin}, \& {Mand
  rini}}]{Baker2017}
{Baker}, D., {Janvier}, M., {D{\'e}moulin}, P., \& {Mand rini}, C.~H. 2017,
  \solphys, 292, 46

\bibitem[{{Baker} {et~al.}(2009){Baker}, {van Driel-Gesztelyi}, {Mandrini},
  {D{\'e}moulin}, \& {Murray}}]{Baker2009}
{Baker}, D., {van Driel-Gesztelyi}, L., {Mandrini}, C.~H., {D{\'e}moulin}, P.,
  \& {Murray}, M.~J. 2009, \apj, 705, 926

\bibitem[{{Boland} {et~al.}(1975){Boland}, {Dyer}, {Firth}, {Gabriel}, {Jones},
  {Jordan}, {McWhirter}, {Monk}, \& {Turner}}]{Boland1975}
{Boland}, B.~C., {Dyer}, E.~P., {Firth}, J.~G., {et~al.} 1975, \mnras, 171, 697

\bibitem[{{Brooks} \& {Warren}(2016)}]{Brooks2016}
{Brooks}, D.~H. \& {Warren}, H.~P. 2016, \apj, 820, 63

\bibitem[{{Brooks} {et~al.}(2020){Brooks}, {Winebarger}, {Savage}, {Warren},
  {De Pontieu}, {Peter}, {Cirtain}, {Golub}, {Kobayashi}, {McIntosh},
  {McKenzie}, {Morton}, {Rachmeler}, {Testa}, {Tiwari}, \&
  {Walsh}}]{Brooks2020}
{Brooks}, D.~H., {Winebarger}, A.~R., {Savage}, S., {et~al.} 2020, \apj, 894,
  144

\bibitem[{{Chae} {et~al.}(1998){Chae}, {Sch{\"u}hle}, \& {Lemaire}}]{Chae1998}
{Chae}, J., {Sch{\"u}hle}, U., \& {Lemaire}, P. 1998, \apj, 505, 957

\bibitem[{{Chen} {et~al.}(2011){Chen}, {Ding}, {Chen}, \& {Harra}}]{Chen2011}
{Chen}, F., {Ding}, M.~D., {Chen}, P.~F., \& {Harra}, L.~K. 2011, \apj, 740,
  116

\bibitem[{{Coyner} \& {Davila}(2011)}]{Coyner2011}
{Coyner}, A.~J. \& {Davila}, J.~M. 2011, \apj, 742, 115

\bibitem[{{Culhane} {et~al.}(2007){Culhane}, {Harra}, {James}, {Al-Janabi},
  {Bradley}, {Chaudry}, {Rees}, {Tandy}, {Thomas}, {Whillock}, {Winter},
  {Doschek}, {Korendyke}, {Brown}, {Myers}, {Mariska}, {Seely}, {Lang}, {Kent},
  {Shaughnessy}, {Young}, {Simnett}, {Castelli}, {Mahmoud}, {Mapson-Menard},
  {Probyn}, {Thomas}, {Davila}, {Dere}, {Windt}, {Shea}, {Hagood}, {Moye},
  {Hara}, {Watanabe}, {Matsuzaki}, {Kosugi}, {Hansteen}, \&
  {Wikstol}}]{Culhane2007SoPh}
{Culhane}, J.~L., {Harra}, L.~K., {James}, A.~M., {et~al.} 2007, \solphys, 243,
  19

\bibitem[{{Dadashi} {et~al.}(2011){Dadashi}, {Teriaca}, \&
  {Solanki}}]{Dadashi2011}
{Dadashi}, N., {Teriaca}, L., \& {Solanki}, S.~K. 2011, \aap, 534, A90

\bibitem[{{De Pontieu} {et~al.}(2017){De Pontieu}, {De Moortel},
  {Martinez-Sykora}, \& {McIntosh}}]{DePontieu2017}
{De Pontieu}, B., {De Moortel}, I., {Martinez-Sykora}, J., \& {McIntosh}, S.~W.
  2017, \apjl, 845, L18

\bibitem[{{De Pontieu} {et~al.}(2009){De Pontieu}, {McIntosh}, {Hansteen}, \&
  {Schrijver}}]{DePontieu2009}
{De Pontieu}, B., {McIntosh}, S.~W., {Hansteen}, V.~H., \& {Schrijver}, C.~J.
  2009, \apjl, 701, L1

\bibitem[{{De Pontieu} {et~al.}(2014){De Pontieu}, {Title}, {Lemen}, {Kushner},
  {Akin}, {Allard}, {Berger}, {Boerner}, {Cheung}, {Chou}, {Drake}, {Duncan},
  {Freeland}, {Heyman}, {Hoffman}, {Hurlburt}, {Lindgren}, {Mathur}, {Rehse},
  {Sabolish}, {Seguin}, {Schrijver}, {Tarbell}, {W{\"u}lser}, {Wolfson},
  {Yanari}, {Mudge}, {Nguyen-Phuc}, {Timmons}, {van Bezooijen}, {Weingrod},
  {Brookner}, {Butcher}, {Dougherty}, {Eder}, {Knagenhjelm}, {Larsen},
  {Mansir}, {Phan}, {Boyle}, {Cheimets}, {DeLuca}, {Golub}, {Gates}, {Hertz},
  {McKillop}, {Park}, {Perry}, {Podgorski}, {Reeves}, {Saar}, {Testa}, {Tian},
  {Weber}, {Dunn}, {Eccles}, {Jaeggli}, {Kankelborg}, {Mashburn}, {Pust},
  {Springer}, {Carvalho}, {Kleint}, {Marmie}, {Mazmanian}, {Pereira}, {Sawyer},
  {Strong}, {Worden}, {Carlsson}, {Hansteen}, {Leenaarts}, {Wiesmann},
  {Aloise}, {Chu}, {Bush}, {Scherrer}, {Brekke}, {Martinez-Sykora}, {Lites},
  {McIntosh}, {Uitenbroek}, {Okamoto}, {Gummin}, {Auker}, {Jerram}, {Pool}, \&
  {Waltham}}]{DePontieu2014}
{De Pontieu}, B., {Title}, A.~M., {Lemen}, J.~R., {et~al.} 2014, \solphys, 289,
  2733

\bibitem[{{Del Zanna}(2008)}]{DelZanna2008}
{Del Zanna}, G. 2008, \aap, 481, L49

\bibitem[{{Del Zanna} {et~al.}(2011){Del Zanna}, {Aulanier}, {Klein}, \&
  {T{\"o}r{\"o}k}}]{DelZanna2011}
{Del Zanna}, G., {Aulanier}, G., {Klein}, K.~L., \& {T{\"o}r{\"o}k}, T. 2011,
  \aap, 526, A137

\bibitem[{{Del Zanna} {et~al.}(2021){Del Zanna}, {Dere}, {Young}, \&
  {Landi}}]{DelZanna2021}
{Del Zanna}, G., {Dere}, K.~P., {Young}, P.~R., \& {Landi}, E. 2021, \apj, 909,
  38

\bibitem[{{Doschek}(2012)}]{Doschek2012}
{Doschek}, G.~A. 2012, \apj, 754, 153

\bibitem[{{Doschek} {et~al.}(2007){Doschek}, {Mariska}, {Warren}, {Brown},
  {Culhane}, {Hara}, {Watanabe}, {Young}, \& {Mason}}]{Doschek2007}
{Doschek}, G.~A., {Mariska}, J.~T., {Warren}, H.~P., {et~al.} 2007, \apjl, 667,
  L109

\bibitem[{{Doschek} {et~al.}(2008){Doschek}, {Warren}, {Mariska}, {Muglach},
  {Culhane}, {Hara}, \& {Watanabe}}]{Doschek2008}
{Doschek}, G.~A., {Warren}, H.~P., {Mariska}, J.~T., {et~al.} 2008, \apj, 686,
  1362

\bibitem[{{Edwards} {et~al.}(2016){Edwards}, {Parnell}, {Harra}, {Culhane}, \&
  {Brooks}}]{Edwards2016}
{Edwards}, S.~J., {Parnell}, C.~E., {Harra}, L.~K., {Culhane}, J.~L., \&
  {Brooks}, D.~H. 2016, \solphys, 291, 117

\bibitem[{{Einaudi} {et~al.}(1999){Einaudi}, {Boncinelli}, {Dahlburg}, \&
  {Karpen}}]{Einaudi1999}
{Einaudi}, G., {Boncinelli}, P., {Dahlburg}, R.~B., \& {Karpen}, J.~T. 1999,
  \jgr, 104, 521

\bibitem[{{Fisk} \& {Schwadron}(2001)}]{Fisk2001}
{Fisk}, L.~A. \& {Schwadron}, N.~A. 2001, \ssr, 97, 21

\bibitem[{{Gordovskyy} {et~al.}(2016){Gordovskyy}, {Kontar}, \&
  {Browning}}]{Gordovskyy2016}
{Gordovskyy}, M., {Kontar}, E.~P., \& {Browning}, P.~K. 2016, \aap, 589, A104

\bibitem[{{Hannah} \& {Kontar}(2012)}]{Hannah2012}
{Hannah}, I.~G. \& {Kontar}, E.~P. 2012, \aap, 539, A146

\bibitem[{{Hara} {et~al.}(2008){Hara}, {Watanabe}, {Harra}, {Culhane}, {Young},
  {Mariska}, \& {Doschek}}]{Hara2008}
{Hara}, H., {Watanabe}, T., {Harra}, L.~K., {et~al.} 2008, \apjl, 678, L67

\bibitem[{{Harra} {et~al.}(2008){Harra}, {Sakao}, {Mandrini}, {Hara}, {Imada},
  {Young}, {van Driel-Gesztelyi}, \& {Baker}}]{Harra2008}
{Harra}, L.~K., {Sakao}, T., {Mandrini}, C.~H., {et~al.} 2008, \apjl, 676, L147

\bibitem[{{Harra} {et~al.}(2017){Harra}, {Ugarte-Urra}, {De Rosa}, {Mand rini},
  {van Driel-Gesztelyi}, {Baker}, {Culhane}, \& {D{\'e}moulin}}]{Harra2017}
{Harra}, L.~K., {Ugarte-Urra}, I., {De Rosa}, M., {et~al.} 2017, \pasj, 69, 47

\bibitem[{{He} {et~al.}(2010){He}, {Marsch}, {Tu}, {Guo}, \& {Tian}}]{He2010}
{He}, J.~S., {Marsch}, E., {Tu}, C.~Y., {Guo}, L.~J., \& {Tian}, H. 2010, \aap,
  516, A14

\bibitem[{{Kojima} {et~al.}(1999){Kojima}, {Fujiki}, {Ohmi}, {Tokumaru},
  {Yokobe}, \& {Hakamada}}]{Kojima1999}
{Kojima}, M., {Fujiki}, K., {Ohmi}, T., {et~al.} 1999, \jgr, 104, 16993

\bibitem[{{Lemen} {et~al.}(2012){Lemen}, {Title}, {Akin}, {Boerner}, {Chou},
  {Drake}, {Duncan}, {Edwards}, {Friedlaender}, {Heyman}, {Hurlburt}, {Katz},
  {Kushner}, {Levay}, {Lindgren}, {Mathur}, {McFeaters}, {Mitchell}, {Rehse},
  {Schrijver}, {Springer}, {Stern}, {Tarbell}, {Wuelser}, {Wolfson}, {Yanari},
  {Bookbinder}, {Cheimets}, {Caldwell}, {Deluca}, {Gates}, {Golub}, {Park},
  {Podgorski}, {Bush}, {Scherrer}, {Gummin}, {Smith}, {Auker}, {Jerram},
  {Pool}, {Soufli}, {Windt}, {Beardsley}, {Clapp}, {Lang}, \&
  {Waltham}}]{Lemen2012}
{Lemen}, J.~R., {Title}, A.~M., {Akin}, D.~J., {et~al.} 2012, \solphys, 275, 17

\bibitem[{MacQueen(1967)}]{macqueen1967}
MacQueen, J. 1967, in Proceedings of the Fifth Berkeley Symposium on
  Mathematical Statistics and Probability, Volume 1: Statistics (Berkeley,
  Calif.: University of California Press), 281--297

\bibitem[{{Marsch} {et~al.}(2008){Marsch}, {Tian}, {Sun}, {Curdt}, \&
  {Wiegelmann}}]{Marsch2008}
{Marsch}, E., {Tian}, H., {Sun}, J., {Curdt}, W., \& {Wiegelmann}, T. 2008,
  \apj, 685, 1262

\bibitem[{{McIntosh} {et~al.}(2012){McIntosh}, {Tian}, {Sechler}, \& {De
  Pontieu}}]{McIntosh2012}
{McIntosh}, S.~W., {Tian}, H., {Sechler}, M., \& {De Pontieu}, B. 2012, \apj,
  749, 60

\bibitem[{{M{\"u}ller} {et~al.}(2013){M{\"u}ller}, {Marsden}, {St. Cyr}, \&
  {Gilbert}}]{Muller2013}
{M{\"u}ller}, D., {Marsden}, R.~G., {St. Cyr}, O.~C., \& {Gilbert}, H.~R. 2013,
  \solphys, 285, 25

\bibitem[{{M{\"u}ller} {et~al.}(2020){M{\"u}ller}, {St. Cyr}, {Zouganelis},
  {Gilbert}, {Marsden}, {Nieves-Chinchilla}, {Antonucci}, {Auch{\`e}re},
  {Berghmans}, {Horbury}, {Howard}, {Krucker}, {Maksimovic}, {Owen}, {Rochus},
  {Rodriguez-Pacheco}, {Romoli}, {Solanki}, {Bruno}, {Carlsson}, {Fludra},
  {Harra}, {Hassler}, {Livi}, {Louarn}, {Peter}, {Sch{\"u}hle}, {Teriaca}, {del
  Toro Iniesta}, {Wimmer-Schweingruber}, {Marsch}, {Velli}, {De Groof},
  {Walsh}, \& {Williams}}]{Muller2020}
{M{\"u}ller}, D., {St. Cyr}, O.~C., {Zouganelis}, I., {et~al.} 2020, \aap, 642,
  A1

\bibitem[{{Murray} {et~al.}(2010){Murray}, {Baker}, {van Driel-Gesztelyi}, \&
  {Sun}}]{Murray2010}
{Murray}, M.~J., {Baker}, D., {van Driel-Gesztelyi}, L., \& {Sun}, J. 2010,
  \solphys, 261, 253

\bibitem[{{Nishizuka} \& {Hara}(2011)}]{Nishizuka2011}
{Nishizuka}, N. \& {Hara}, H. 2011, \apjl, 737, L43

\bibitem[{{Panos} {et~al.}(2018){Panos}, {Kleint}, {Huwyler}, {Krucker},
  {Melchior}, {Ullmann}, \& {Voloshynovskiy}}]{Panos2018}
{Panos}, B., {Kleint}, L., {Huwyler}, C., {et~al.} 2018, \apj, 861, 62

\bibitem[{{Parker}(1988)}]{Parker1988}
{Parker}, E.~N. 1988, \apj, 330, 474

\bibitem[{{Patsourakos} \& {Klimchuk}(2006)}]{Patsourakos2006}
{Patsourakos}, S. \& {Klimchuk}, J.~A. 2006, \apj, 647, 1452

\bibitem[{{Pereira} {et~al.}(2013){Pereira}, {Leenaarts}, {De Pontieu},
  {Carlsson}, \& {Uitenbroek}}]{Pereira2013}
{Pereira}, T.~M.~D., {Leenaarts}, J., {De Pontieu}, B., {Carlsson}, M., \&
  {Uitenbroek}, H. 2013, \apj, 778, 143

\bibitem[{{Peter}(2001)}]{Peter2001}
{Peter}, H. 2001, \aap, 374, 1108

\bibitem[{{Peter} \& {Judge}(1999)}]{Peter1999}
{Peter}, H. \& {Judge}, P.~G. 1999, \apj, 522, 1148

\bibitem[{{Polito} {et~al.}(2020){Polito}, {De Pontieu}, {Testa}, {Brooks}, \&
  {Hansteen}}]{Polito2020}
{Polito}, V., {De Pontieu}, B., {Testa}, P., {Brooks}, D.~H., \& {Hansteen}, V.
  2020, \apj, 903, 68

\bibitem[{{Rappazzo} {et~al.}(2005){Rappazzo}, {Velli}, {Einaudi}, \&
  {Dahlburg}}]{Rappazzo2005}
{Rappazzo}, A.~F., {Velli}, M., {Einaudi}, G., \& {Dahlburg}, R.~B. 2005, \apj,
  633, 474

\bibitem[{{R{\'e}gnier} {et~al.}(2008){R{\'e}gnier}, {Priest}, \&
  {Hood}}]{Regnier2008}
{R{\'e}gnier}, S., {Priest}, E.~R., \& {Hood}, A.~W. 2008, \aap, 491, 297

\bibitem[{{Sakao} {et~al.}(2007){Sakao}, {Kano}, {Narukage}, {Kotoku}, {Bando},
  {DeLuca}, {Lundquist}, {Tsuneta}, {Harra}, {Katsukawa}, {Kubo}, {Hara},
  {Matsuzaki}, {Shimojo}, {Bookbinder}, {Golub}, {Korreck}, {Su}, {Shibasaki},
  {Shimizu}, \& {Nakatani}}]{Sakao2007}
{Sakao}, T., {Kano}, R., {Narukage}, N., {et~al.} 2007, Science, 318, 1585

\bibitem[{{Spice Consortium} {et~al.}(2020){Spice Consortium}, {Anderson},
  {Appourchaux}, {Auch{\`e}re}, {Aznar Cuadrado}, {Barbay}, {Baudin},
  {Beardsley}, {Bocchialini}, {Borgo}, {Bruzzi}, {Buchlin}, {Burton},
  {B{\"u}chel}, {Caldwell}, {Caminade}, {Carlsson}, {Curdt}, {Davenne},
  {Davila}, {Deforest}, {Del Zanna}, {Drummond}, {Dubau}, {Dumesnil}, {Dunn},
  {Eccleston}, {Fludra}, {Fredvik}, {Gabriel}, {Giunta}, {Gottwald}, {Griffin},
  {Grundy}, {Guest}, {Gyo}, {Haberreiter}, {Hansteen}, {Harrison}, {Hassler},
  {Haugan}, {Howe}, {Janvier}, {Klein}, {Koller}, {Kucera}, {Kouliche},
  {Marsch}, {Marshall}, {Marshall}, {Matthews}, {McQuirk}, {Meining},
  {Mercier}, {Morris}, {Morse}, {Munro}, {Parenti}, {Pastor-Santos}, {Peter},
  {Pfiffner}, {Phelan}, {Philippon}, {Richards}, {Rogers}, {Sawyer},
  {Schlatter}, {Schmutz}, {Sch{\"u}hle}, {Shaughnessy}, {Sidher}, {Solanki},
  {Speight}, {Spescha}, {Szwec}, {Tamiatto}, {Teriaca}, {Thompson}, {Tosh},
  {Tustain}, {Vial}, {Walls}, {Waltham}, {Wimmer-Schweingruber}, {Woodward},
  {Young}, {de Groof}, {Pacros}, {Williams}, \& {M{\"u}ller}}]{Spice2020}
{Spice Consortium}, {Anderson}, M., {Appourchaux}, T., {et~al.} 2020, \aap,
  642, A14

\bibitem[{{Teriaca} {et~al.}(1999){Teriaca}, {Banerjee}, \&
  {Doyle}}]{Teriaca1999}
{Teriaca}, L., {Banerjee}, D., \& {Doyle}, J.~G. 1999, \aap, 349, 636

\bibitem[{{Thorndike}(1953)}]{Thorndike1953}
{Thorndike}, R.~L. 1953, Psychometrika, 18, 62

\bibitem[{{Tian} {et~al.}(2009){Tian}, {Marsch}, {Curdt}, \& {He}}]{Tian2009}
{Tian}, H., {Marsch}, E., {Curdt}, W., \& {He}, J. 2009, \apj, 704, 883

\bibitem[{{Tian} {et~al.}(2011){Tian}, {McIntosh}, {De Pontieu},
  {Mart{\'\i}nez-Sykora}, {Sechler}, \& {Wang}}]{Tian2011}
{Tian}, H., {McIntosh}, S.~W., {De Pontieu}, B., {et~al.} 2011, \apj, 738, 18

\bibitem[{{Tripathi} {et~al.}(2009){Tripathi}, {Mason}, {Dwivedi}, {del Zanna},
  \& {Young}}]{Tripathi2009}
{Tripathi}, D., {Mason}, H.~E., {Dwivedi}, B.~N., {del Zanna}, G., \& {Young},
  P.~R. 2009, \apj, 694, 1256

\bibitem[{{Ugarte-Urra} \& {Warren}(2011)}]{UgarteUrra2011}
{Ugarte-Urra}, I. \& {Warren}, H.~P. 2011, \apj, 730, 37

\bibitem[{{van Driel-Gesztelyi} {et~al.}(2012){van Driel-Gesztelyi}, {Culhane},
  {Baker}, {D{\'e}moulin}, {Mandrini}, {DeRosa}, {Rouillard}, {Opitz},
  {Stenborg}, {Vourlidas}, \& {Brooks}}]{vanDriel2012}
{van Driel-Gesztelyi}, L., {Culhane}, J.~L., {Baker}, D., {et~al.} 2012,
  \solphys, 281, 237

\bibitem[{{Wang} {et~al.}(2017){Wang}, {Sim{\~o}es}, {Jeffrey}, {Fletcher},
  {Wright}, \& {Hannah}}]{Wang2017}
{Wang}, J., {Sim{\~o}es}, P.~J.~A., {Jeffrey}, N.~L.~S., {et~al.} 2017, \apjl,
  847, L1

\bibitem[{{Wang} {et~al.}(2015){Wang}, {Li}, {Ma}, {Zhang}, \&
  {Lee}}]{Wang2015}
{Wang}, L.~C., {Li}, L.~J., {Ma}, Z.~W., {Zhang}, X., \& {Lee}, L.~C. 2015,
  Physics Letters A, 379, 2068

\bibitem[{{Wang} {et~al.}(2013){Wang}, {McIntosh}, {Curdt}, {Tian}, {Peter}, \&
  {Xia}}]{Wang2013}
{Wang}, X., {McIntosh}, S.~W., {Curdt}, W., {et~al.} 2013, \aap, 557, A126

\bibitem[{{Wang}(1994)}]{Wang1994}
{Wang}, Y.~M. 1994, \apjl, 437, L67

\bibitem[{{Wang} \& {Sheeley}(1990)}]{Wang1990b}
{Wang}, Y.~M. \& {Sheeley}, N.~R., J. 1990, \apj, 355, 726

\bibitem[{{Wang} {et~al.}(1990){Wang}, {Sheeley}, \& {Nash}}]{Wang1990}
{Wang}, Y.~M., {Sheeley}, N.~R., J., \& {Nash}, A.~G. 1990, \nat, 347, 439

\bibitem[{{Wang} {et~al.}(2000){Wang}, {Sheeley}, {Socker}, {Howard}, \&
  {Rich}}]{Wang2000}
{Wang}, Y.~M., {Sheeley}, N.~R., {Socker}, D.~G., {Howard}, R.~A., \& {Rich},
  N.~B. 2000, \jgr, 105, 25133

\bibitem[{{Warren}(2006)}]{Warren2006}
{Warren}, H.~P. 2006, \apj, 637, 522

\bibitem[{{Warren} {et~al.}(2014){Warren}, {Ugarte-Urra}, \&
  {Landi}}]{Warren2014}
{Warren}, H.~P., {Ugarte-Urra}, I., \& {Landi}, E. 2014, \apjs, 213, 11

\bibitem[{{Warren} {et~al.}(2011){Warren}, {Ugarte-Urra}, {Young}, \&
  {Stenborg}}]{Warren2011}
{Warren}, H.~P., {Ugarte-Urra}, I., {Young}, P.~R., \& {Stenborg}, G. 2011,
  \apj, 727, 58

\bibitem[{{Warren} {et~al.}(2012){Warren}, {Winebarger}, \&
  {Brooks}}]{Warren2012}
{Warren}, H.~P., {Winebarger}, A.~R., \& {Brooks}, D.~H. 2012, \apj, 759, 141

\bibitem[{{Zangrilli} \& {Poletto}(2016)}]{Zangrilli2016}
{Zangrilli}, L. \& {Poletto}, G. 2016, \aap, 594, A40

\end{thebibliography}
\end{document}